\documentclass[Numbered,pdflatex]{sn-jnl}
\usepackage{amsfonts}
\usepackage{newlfont}
\usepackage{stmaryrd}
\usepackage{orcidlink}

\usepackage{amsmath}
\DeclareMathOperator{\arcsinh}{arcsinh}
\DeclareMathOperator{\arccosh}{arccosh}
\usepackage{anyfontsize}
\usepackage{amscd}
\let\ve=\varepsilon
\newcommand{\id}{\textrm{d}}
\SetSymbolFont{stmry}{bold}{U}{stmry}{m}{n}

\let\oldsqrt\sqrt
\def\sqrt{\mathpalette\DHLhksqrt}
\def\DHLhksqrt#1#2{%
	\setbox0=\hbox{$#1\oldsqrt{#2\,}$}\dimen0=\ht0
	\advance\dimen0-0.2\ht0
	\setbox2=\hbox{\vrule height\ht0 depth -\dimen0}%
	{\box0\lower0.4pt\box2}}
\usepackage{lmodern}
\usepackage{titlesec}

\begin{document}

\title{Noether's theorem applied to \texttt{GENERIC}}
\author*[1]{Aaron Beyen \orcidlink{0000-0002-4341-7661}}\email{aaron.beyen@kuleuven.be}

\author[1]{Christian Maes \orcidlink{0000-0002-0188-697X}}
\affil[1]{Department of Physics and Astronomy, KU Leuven\\ 3001 Leuven, Belgium}

\abstract{The last decades have seen growing interest in connecting principles of thermodynamics with methods from analytical mechanics.  The thermodynamic formalism has become an inspiring framework in the study of smooth dynamical systems, and pioneering works of Helmholtz, Clausius, and Boltzmann have been reinstated as possible dynamical foundations of the (first part of the) Heat Theorem.  The present paper follows the work of Wald \textit{et al.}, where black hole entropy was identified as a Noether charge. The adiabatic invariance of the thermodynamic entropy indeed invites a connection with Noether's theorem, and has been the subject of various papers. Here we add the case of \texttt{GENERIC}, a macroscopic dynamics whose acronym stands for ``\textit{General Equation for Non-Equilibrium Reversible-Irreversible Coupling}''. Its evolution has two contributions: a dissipative part, which is of a generalized gradient descent form, and a Hamiltonian flow.  We consider a quasistatic protocol for external parameters, and we embed \texttt{GENERIC} as the zero-cost flow for a Lagrangian governing the dynamical fluctuations.  We find a continuous symmetry of the corresponding path-space action with the thermodynamic entropy as Noether charge, both in the Lagrangian and Hamiltonian formalisms.  We make the calculations explicit through the example of an inertial probe with nonlinear friction.}

\keywords{\texttt{GENERIC}, Noether's theorem, Entropy invariance, Dynamical fluctuations}

\maketitle

\section{Introduction}
The thermodynamic entropy $S$ is defined by recognizing that ``reversible heat over temperature'' is an exact differential
\begin{equation}\label{ds}
 \frac{\delta Q^\text{rev}}{T} = \id S 
\end{equation}
That result is traditionally called the first part of the Clausius Heat Theorem, \cite{clausius1865}.  It implies that $\id S=0$ for $\delta Q^\text{rev} =0$, which is abbreviated by saying that the entropy is an adiabatic invariant.  For a more precise understanding, it is important to keep in mind that we are dealing with quasistatic transformations of parameters (where, for instance, temperature $T$ or the energy function is changing very slowly) imposed on a reversible dynamics. The latter means that a detailed balance condition is verified by the dynamics when the parameters are held fixed.  Such a setup with its corresponding assumptions is useful for a more precise and mesoscopic derivation of \eqref{ds} and the ensuing adiabatic invariance, \cite{ruelle,jonalasinio2023clausius, Saito_2011, prigo, heatconduction, Komatsu_2010, Bertini_2012, clausiusstationarystates, Maes_2015}.  The present paper (as previous ones like \cite{Sasa_original, mecos}) results from the wish to connect this principle of invariance of the thermodynamic entropy for reversible evolutions of thermally isolated systems with the Noether theorem of analytical mechanics, \cite{Noether1918}. \\

A previous paper \cite{Beyengradientflow} has extended a result of Sasa {\it et al.} \cite{langevin_noether} about (Onsager-type) Langevin dynamics to more general (nonlinear) gradient flow dynamics \cite{genericmaes, ottinger2005beyond}.  By introducing the appropriate symmetry transformations, the entropy could be interpreted as a Noether charge, both in the Lagrangian and Hamiltonian formalism. Differently, the papers \cite{Sasa_original, mecos} start from a microscopic, mechanical setup and show that the (phase space) entropy can also be derived as a Noether charge. Interestingly, Noether's theorem has also appeared in the machine learning community to describe conserved quantities along gradient descent and gradient flow \cite{tanaka2021noetherslearningdynamicsrole, zhao2023symmetriesflatminimaconserved}.   \\

Our main result here is a derivation of the adiabatic invariance of entropy from a Noether theorem applied to {\tt GENERIC}, \cite{genericmaes, ottinger2005beyond, PavelkaKlikaGrmela+2018, Grmela_2018}, which can be viewed as a continuation and extension of \cite{Beyengradientflow}. {\tt GENERIC} is the acronym for ``The Generalized Equation for Non-Equilibrium Reversible-Irreversible Coupling''  whose central structure was originally formulated in \cite{DZYALOSHINSKII198067, Grmela1984, MORRISON1984423, KAUFMAN1984419} (although the equations are not yet called \texttt{GENERIC} in these papers) and finalized in \cite{complexfluids1, complexfluids2} (where the acronym first appeared). In its simplest form, it gives  the structure of a class of evolution equations describing the return to equilibrium, where, besides the dissipative gradient part in the equation, there is also a Hamiltonian/reversible flow creating a stationary current:
\begin{equation}\label{genz}
\dot{z} = \underbrace{L(z)\ \id E(z)}_{\text{reversible part}} +\underbrace{M(z)\ \id S (z)}_{\text{dissipative part}}
\end{equation}
Here, $z=z(t)$ is the thermodynamic or macroscopic state (\textit{e.g.} particle, momentum, and energy density), $E$ the total energy, $S$ the total entropy, and the ``$\id$'' refers to a (functional) derivative. The matrix $L$ is antisymmetric and the matrix $M$ is positive-definite, with
\begin{align} \label{M dE = 0}
L(z) \ \id S(z) = 0 = M(z) \ \id E(z) 
\end{align}
so that the energy is conserved, $\dot{E} = 0$,  and the entropy is nondecreasing,  $\dot{S} \geq 0$,  \cite{ottinger2005beyond, complexfluids1, complexfluids2}. A reassuring feature of the \texttt{GENERIC} framework is that there exists a link to statistical mechanics. By eliminating the fast degrees of freedom, the projection-operator formalism \cite{zwanzig, zwanzig2001nonequilibrium, bouchard2007morizwanzigequationstimedependentliouvillian} produces equations of the \texttt{GENERIC} form; see \cite{ottinger2005beyond, ottingerprojection1, ottingerprojection2, ottingerprojection3}. \\
Alternatively, as introduced by \cite{Grmela1984,  MORRISON1984423, KAUFMAN1984419},  Eq. \eqref{genz} can be rewritten in a bracket representation 
 \begin{align}
  \dot{z} &= \{z, E \} + [z, S] \nonumber \\
     \{A,B\} &= \id A(z)\cdot L(z) \ \id B(z), \qquad [A,B] = \id A(z) \cdot M(z) \ \id B(z) \label{brackets}
 \end{align}
with Poisson bracket $\{\cdot , \cdot \}$ associated with the antisymmetric matrix $L(z)$ and a dissipative bracket $[\cdot, \cdot]$ with $M(z)$; see \cite{ottinger2005beyond} for more details. In that sense, \texttt{GENERIC} is a time-evolution on a Poisson manifold for the reversible part, extended with a gradient flow on a metric manifold to include irreversible thermodynamics.\\

As explained in \cite{PavelkaKlikaGrmela+2018, Grmela_2018,complexfluids1, Mielke2011, GRMELA2012976}, the linearity of the dissipative part $[A, B]$ in $\id B $ can be relaxed, allowing for nonlinear dissipation (non-quadratic dissipation potential) in \texttt{GENERIC}, leading to 
\begin{align}
     \dot{z} &= L(z) \ \id E(z) + \partial_f \psi^{\star}\left(f;  z\right)\Big|_{f = D^{\dagger} \id S(z)/2} \label{nonlinear generic}
\end{align}
The dissipation potential $\psi^{\star}(f;z)$ is convex in $f$ for all $z$ and the operator $D$ can be minus a divergence, or the identity operator, or a stoichiometry matrix, \textit{etc.} with adjoint $D^{\dagger}$ defined by the relation $a \cdot D b = b \cdot D^{\dagger} a$. Of course, \eqref{nonlinear generic} reduces to \eqref{genz} for
\begin{align*}
    \psi^{\star}(f;z) =f \cdot X(z) \  f  , \qquad M(z) = D \cdot X(z) \  D^{\dagger}
\end{align*}
for a  symmetric positive semi-definite operator $X = X^T$ (Onsager reciprocity).\\

Examples of \texttt{GENERIC} are numerous in continuum mechanics and thermodynamics. Some notable examples include the Navier-Stokes-Fourier hydrodynamic equations, the Boltzmann kinetic equation and other classical irreversible processes  \cite{ottinger2005beyond, PavelkaKlikaGrmela+2018, complexfluids1, complexfluids2}, non-Fourier heat conduction (like the Maxwell-Cattaneo-Vernotte equation \cite{maxwell1867dynamical, cattaneo1958forme, vernotte1958paradoxes}) \cite{PavelkaKlikaGrmela+2018, Grmela_2019, nonfourier}, chemical kinetics \cite{GRMELA2012976, AJJI2023133642}, the rheology of polymeric and other complex fluids \cite{PavelkaKlikaGrmela+2018, complexfluids1, complexfluids2, grmela2025rheologicalmodelinggenericonsager, WAGNER2001177, twophasemodels}, thermomechanical models of dissipative materials \cite{Mielke2011, hutter2012viscoplastic} and liquid crystals \cite{liquidcrystal1}. \\

\underline{Plan of the paper:} 
The paper opens in Section \ref{section pre generic} by introducing a less constrained (yet, as general) version of the equations \eqref{genz}--\eqref{nonlinear generic}, called pre-{\tt GENERIC}, where a nondecreasing entropy can still be identified.  We focus on the quasistatic limit, where the control parameters vary slowly, and the reference evolution is reversible. Entropy is used to characterize the equilibrium state, where quasistatic reversible trajectories are close to a sequence of thermodynamic states that are in instantaneous equilibrium.\\ Next, in Section \ref{section zero-cost flow}, we explain how the pre-{\tt GENERIC} structure can be embedded in the setup of dynamical fluctuation theory.  Pre-{\tt GENERIC} appears as the zero-cost flow for a Lagrangian specifying a path-space action that itself governs trajectory probabilities.\\ 
The Lagrangian and its properties, as well as the corresponding Hamiltonian formalism, are discussed in Section \ref{section on lagrangian and ham}.\\ 
Section \ref{first time entropy as noether section} contains the main result and explains how, in an adiabatic setup, the entropy becomes a Noether charge, and hence is invariant in the quasistatic limit. This calculation is performed in both the Lagrangian and Hamiltonian formalisms. We understand the entropy as a Noether charge for a continuous symmetry that shifts the thermodynamic current/force, \cite{onsager19311, onsager19312}, under quasistatic, reversible, adiabatic conditions. For non-adiabatic conditions, one uses the appropriate free energy instead of the entropy. A natural connection, therefore, arises between a cornerstone of thermodynamics (Clausius Heat Theorem) and mechanics (Noether's theorem).\\ Finally, much of all that is illustrated in Section \ref{section nonlinear friction first time} with the example of a macroscopic probe subject to nonlinear friction.

\section{Quasistatic protocol on pre-\texttt{GENERIC}}\label{section pre generic}
One of the requirements for the thermodynamic entropy to remain invariant along a transformation is taking a quasistatic limit, where we remain close to equilibrium at each point. We introduce that setup here for the pre-\texttt{GENERIC} equations.

\subsection{Pre-\texttt{GENERIC} dynamics}\label{subsection pre generic dynamics}
We consider a less constrained version of 
\eqref{nonlinear generic}, which can be seen as a precursor to \texttt{GENERIC}, and is called pre-\texttt{GENERIC}, \cite{genericmaes, Kraaij_2020}.  It renounces the
existence of a conserved energy $E$, but we keep a (Hamiltonian) flow in terms of a current $J^H_{z}$ as a function of $z$. More specifically, the considered evolution equation for state $z(t)$ $\in \cal M$, element of a smooth manifold $\cal M$, generalizes \eqref{nonlinear generic} to the form
\begin{align}\label{flow generic}
    \dot{z}(t) & = D j_{z(t)}, \qquad 
    j_z  =  \underbrace{J_z^H}_{\text{Hamiltonian flow}} +  \underbrace{J_z^S}_{\substack{\text{(nonlinear)}\\ \text{dissipative part}}}
\end{align}
where the entropic current has the same form as in \eqref{nonlinear generic}
\begin{align}\label{current flow generic}
    J_z^S = \partial_f \psi^{\star}\left( D^{\dagger} \id S(z)/2 ;  z\right)
\end{align}
We require that $\psi^{\star}(f;z)$ is convex in $f$ for all $z$, 
\begin{align*}
    \psi^{\star}(f; z) \geq \psi^{\star}(0;z) = 0 
\end{align*}
and that $\psi^{\star}$ is symmetric in $\pm f$. Then also $\partial_j \psi^{\star}(0;z) = 0$ and
\begin{align}\label{Js = 0}
   J_z^S = 0 \qquad \text{whenever} \quad D^{\dagger} \id S(z) = 0 
\end{align}
The function $S(z)$ in \eqref{current flow generic} is called the entropy of state $z$. In fact, by requiring the orthogonality 
\begin{equation}\label{orthogonality condition J}
\forall z : \quad     J_z^H \cdot D^{\dagger} \id S(z) = 0
\end{equation}
we get the monotonicity of that entropy $S$ along \eqref{flow generic}, {\it i.e.},
\begin{align}\label{dsdt geq 0}
    \frac{\id }{\id t} \left(S(z(t)) \right) &= \dot{z}(t) \cdot \id S(z(t)) \\
    & = D J^H_{z(t)} \cdot \id S(z(t)) + D \partial_f \psi^{\star}(D^{\dagger} \id S(z(t))/2 ;z(t)) \cdot \id S(z(t)) \nonumber \\
    &=  \partial_f \psi^{\star}(D^{\dagger} \id S(z(t))/2 ;z(t)) \cdot D^{\dagger}\id S(z(t))  \geq 0 \nonumber
\end{align}
where we used the convexity of $\psi^{\star}$ in the last line. When studying open systems, {\it e.g.}, with a surrounding heat bath at constant temperature in which energy is being dissipated, instead of dealing with the total entropy, we take a description in terms of a free energy $\cal F$ (or some other thermodynamic potential) and then we need to replace minus $S$ by $\cal F$ in the equations. Thus, for closed, isolated systems (adiabatic condition), we show that entropy is the Noether charge, while for open systems with heat to the environment, the role of entropy gets replaced by the appropriate free energy. That is an important distinction to keep in mind. \\

This manuscript opts for pre-\texttt{GENERIC} over \texttt{GENERIC} for two primary reasons. First, \texttt{GENERIC} often requires adding extra variables (\textit{e.g.} an internal energy $e$) to complete the structure \eqref{genz}, which are not always directly or physically given. Indeed, many examples, \textit{e.g.} the one in Section \ref{section nonlinear friction first time} and others in \cite{genericmaes}, fall outside the \texttt{GENERIC} framework. Second, as shown in \cite{Kraaij_2020}, pre-\texttt{GENERIC} formally implies \texttt{GENERIC} and is thus the more general setup. To obtain the full \texttt{GENERIC} structure, one should add an auxiliary scalar variable $E$ to
fix the two conditions, $M(z) \ \id E = 0$ \eqref{M dE = 0} and $\dot{E} = 0$, which were not specified in pre-\texttt{GENERIC}.

\subsection{Quasistatic analysis}\label{qas}
The currents $j_z, J_z^H, J_z^S$ in \eqref{current flow generic} are functions of $z$ that can also depend on external parameters $\lambda$ such as external pressure, temperature, or coupling constants, but in what follows we suppress their $\lambda-$dependence for notational convenience. See also the example in Section \ref{section nonlinear friction first time}. 
Such parameter dependence typically arises because the relaxation of $z$ depends on energy and entropy functions that may depend on $\lambda$, and the currents are driven by thermodynamic forces directly derived from these functions. \\
We consider an external protocol $\lambda^{(\varepsilon)}(t) = \lambda(\varepsilon t)$ for times $t \in [t_1 = \tau_1/\ve, t_2 = \tau_2/\varepsilon]$, where the rate $\varepsilon \downarrow 0$ in the quasistatic limit. 
We assume that the protocol enters the entropy $S=S_{\lambda(\ve t)}(z)$,  which becomes a time-dependent function of the state $z$.  Then, along any possible trajectory $z(t)$, we find
\begin{align}\label{zdot ds}
\frac{\id}{\id t}\left(S_{\lambda(\ve t)}(z(t)) \right) &= \dot{z}(t) \cdot \id S_{\lambda(\ve t)}(z(t)) + \ve  \dot{\lambda}(\ve t) \cdot \partial_\lambda S_{\lambda(\ve t)}(z(t))
\end{align}
For the quasistatic protocol $\lambda(\ve t)$, instead of working with a general trajectory $z(t)$, we focus on a class of ``quasistatic trajectories'' which we define to be of the form
  \begin{align}\label{zbar}
    \bar{z}(t) & = z^{*}_{\lambda(\varepsilon t)} + \varepsilon \,\cal Z(\ve t)  + O(\varepsilon^2)
\end{align}
where $z^{*}_{\lambda}$ is assumed to be the unique state that satisfies
\begin{equation}\label{fixed point z dS}
  \partial_\lambda S_{\lambda}(z_{\lambda}^*) = 0, \quad \id S_\lambda(z^*_\lambda) = 0, \quad J^H_{z_{\lambda}^*}  = 0
\end{equation}
The first two conditions \eqref{fixed point z dS} mean to imply that $z_\lambda^*$ represents the equilibrium state, maximizing the entropy and inducing a vanishing thermodynamic force $\partial_\lambda S_\lambda$. Therefore, the trajectories \eqref{zbar} are instantaneously in equilibrium with some (unspecified) correction $\cal Z(\ve t)$ of order $O(\ve)$. Eq. \eqref{zbar} should thus be seen as an expansion around the equilibrium value, not a strict expansion in $\ve$. \\
They are called quasistatic trajectories since they depend, up to order $O(\ve^1)$, only on the slow time variable $\tau = \ve t$ which is `slow with respect to the relevant relaxation times.' Since the state $z$ can contain velocity degrees of freedom (or other kinetic variables), \textit{e.g.}, as shown in Section \ref{section nonlinear friction first time}, slow here implies that the acceleration and inertial effects become less relevant (even though they are present). That is related to the third condition in \eqref{fixed point z dS}.  Indeed, the orthogonality condition \eqref{orthogonality condition J} for $z \to \bar{z}$ also implies
\begin{align*}
    0 = J_{\bar{z}}^H \cdot D^{\dagger} \id S(\bar{z}) &= J_{z_{\lambda(\ve t)}^*} \cdot D^{\dagger} \id S_{\lambda(\ve t)}(z_{\lambda(\ve t)}^*) + \ve \Bigg[\cal Z(\ve t) \ \id J_{z_{\lambda(\ve t)}^*}^H \cdot D^{\dagger} \id S_{\lambda(\ve t)}(z_{\lambda(\ve t)}^*) \\
    & \hspace{3 cm} + J^H_{z_{\lambda(\ve t)}^*} \cdot D^{\dagger} \left(\cal Z(\ve t) \ \id^2 S_{\lambda(\ve t)}(z_{\lambda(\ve t)}^*) \right)  \Bigg] + O(\ve^2) \\
    & = \ve  J^H_{z_{\lambda(\ve t)}^*} \cdot D^{\dagger} \left(\cal Z(\ve t) \ \id^2 S_{\lambda(\ve t)}(z_{\lambda(\ve t)}^*) \right) + O(\ve^2)
\end{align*}
where we have used the first two conditions of \eqref{fixed point z dS}. Since \eqref{orthogonality condition J} holds for all $\bar{z}(t)$ and thus for every $\cal Z(\ve t)$ in $\bar{z}$, it follows that $J_z^H$ must satisfy 
the third condition in \eqref{fixed point z dS}, $J^H_{z_{\lambda}^*}  = 0$,  for consistency with the quasistatic trajectories \eqref{zbar}. It means there are no dissipationless oscillations and the dissipative term $J_z^S$ dominates, even though there can be a Hamiltonian/inertial flow $J_z^H$. \\
As a consequence, looking back at \eqref{zdot ds}, $\bar{z}(t)$ satisfies $\frac{\id}{\id t}\left(S_{\lambda(\ve t)}(\bar{z}(t)) \right) = O(\ve^2)$ and hence \newpage  
 \begin{align}
  \lim_{\ve \downarrow 0} \left( S_{\lambda(\ve t_2)}(\bar{z}(t_2)) - S_{\lambda(\ve(t_1))}(\bar{z}(t_1)) \right) &=   \lim_{\ve \downarrow 0}\int_{t_1}^{t_2} \id t \ \frac{\id}{\id t}\left(S_{\lambda(\ve t)}(\bar{z}(t)) \right) \nonumber \\
  & =  \lim_{\ve \downarrow 0}\int_{t_1}^{t_2} \id t \ O(\ve^2) =  \lim_{\ve \downarrow 0} O(\ve) = 0 \label{delta s = 0}
 \end{align}
for any reasonable choice of ${\cal Z}(\ve t)$ in \eqref{zbar} keeping the $O(\ve^2)$ terms bounded when integrating. In other words, correct to quadratic order in $\ve$, the entropy is invariant over the trajectories $\bar{z}(t)$; the change of entropy over a time-span of order $\ve^{-1}$ is of order $\ve$, and hence vanishes in the quasistatic limit. The main result of the present paper is the derivation from Noether's theorem of that well-known invariance \eqref{delta s = 0}. \\

Of course, the most important example of how \eqref{zbar} arises, is the solution of the pre-\texttt{GENERIC} dynamics \eqref{flow generic} with quasistatic control parameter $\lambda(\ve t)$, which we denote as $\dot{z}^{(\ve)}(t)$ and satisfies
\begin{align}
  &\dot{z}^{(\ve)}(t) = D j_{z^{(\ve)}(t)} =  D\left[  J^H_{z^{(\ve)}(t)}+ \partial_f \psi^\star\left( \frac{1}{2} D^\dagger \id S_{\lambda(\varepsilon t)}(z^{(\ve)}(t));z^{(\ve)}(t) \right) \right]\label{quasistatic gradient flow eq}
\end{align}
Assuming we start from $z_\lambda^*$, its solution follows the quasistatic expansion
\begin{align}
& z^{(\ve)}(t)  = z^{*}_{\lambda(\varepsilon t)} + \varepsilon \,\zeta(\ve t)  + O(\varepsilon^2) \label{expa}
\end{align}
for which \eqref{quasistatic gradient flow eq} becomes
\begin{align*}
 &\ve \dot{\lambda}(\ve t) \ \partial_{\lambda} z_{\lambda(\ve t)}^* 
    + O(\ve^2) = D J_{z_{\lambda(\ve t)}^*}^H + \ve  D \left[ \zeta(\ve t) \ \left(\id J_{z_{\lambda(\ve t)}^*}^H + \id J^S_{z_{\lambda(\ve t)}^*} \right) \right]  + O(\ve^2) \nonumber
\end{align*}
As in \eqref{zbar}, Eq. \eqref{expa} should be seen as an expansion around the instantaneous equilibrium value and not a strict expansion in $\ve$. We used $\zeta(\ve t)$ (instead of $\cal Z(\ve t)$) to indicate the specific trajectory in the class \eqref{zbar} which also solves the pre-\texttt{GENERIC} dynamics. Comparing order by order in $\ve$ and using \eqref{Js = 0},  \eqref{fixed point z dS}, it follows that 
\begin{align}
    O(\ve^0) &: \qquad 0 = D J_{z_{\lambda(\ve t)}^*}^H \label{current J^H = 0} \\
    O(\ve^1) &: \qquad \dot{\lambda}(\ve t) \ \partial_{\lambda} z_{\lambda(\ve t)}^* = D \left[ \zeta(\ve t) \ \left(\id J_{z_{\lambda(\ve t)}^*}^H + \id J^S_{z_{\lambda(\ve t)}^*} \right) \right]  \label{equation cal Z}
\end{align}
As we have seen, \eqref{current J^H = 0} is already a consequence of the orthogonality \eqref{orthogonality condition J}. It further implies
\begin{align}\label{j^* = 0}
  \dot{z}_\lambda^* = D j_{z^*_\lambda}=0  
\end{align}
as expected for an equilibrium solution. 
At the next order, \eqref{equation cal Z} yields an equation for $\zeta(\ve t)$ with formal solution
\begin{align*}
    \zeta(\ve t) = \frac{D^{-1} \left(\dot{\lambda}(\ve t) \ \partial_{\lambda} z_{\lambda(\ve t)}^* \right)}{\id J_{z_{\lambda(\ve t)}^*}^H + \id J^S_{z_{\lambda(\ve t)}^*}}
\end{align*}
for some inverse operator $D^{-1}$ and assuming the denominator does not vanish. \\

In the next section, we introduce the action used to apply Noether's theorem for pre-\texttt{GENERIC}.  For that purpose, it is already important to note here the difference between $\bar{z}(t)$ and $z^{(\ve)}(t)$. In the context of Noether's theorem and the corresponding transformations of the action, one first focuses on general trajectories, without requiring that they solve the equations of motion and hence be physical solutions. Only after showing there is a (quasi)symmetry of this action, \cite{quasisymmetry}, does one impose the equations of motion to show that the Noether charge is conserved. Likewise, we first focus on the general class of trajectories \eqref{zbar} in Noether's theorem, after which we impose the pre-\texttt{GENERIC} dynamics to obtain the conservation of $S_\lambda$ for the physical trajectory \eqref{quasistatic gradient flow eq}. 

\section{Pre-\texttt{GENERIC} as zero-cost flow}\label{section zero-cost flow}
Before we embark with time-dependent protocols $\lambda(\ve t)$ as in and below \eqref{zdot ds}, we wish to be clear about the main setup allowing the derivation of \eqref{delta s = 0} from Noether's theorem.  It is based on the embedding of the pre-\texttt{GENERIC} \eqref{flow generic} into the probabilistic framework of dynamical fluctuation theory, which we first expose in the case without (time-dependent) control parameters $\lambda$. The protocol will be reinstated in Section \ref{first time entropy as noether section}.

\subsection{Lagrangian formalism}
We take the point of view that \eqref{flow generic} generates trajectories that can be characterized as the ``typical'' ones within a larger set of possible trajectories $\gamma = \{z(t), j(t) \}_{t_1 \leq t \leq t_2} $ of states and currents over some time-interval $[t_1,t_2]$ that all satisfy $\dot z(t) = D\,j(t)$. To realize that, we need the larger framework of path-probabilities (giving the probabilities of trajectories $\gamma$), which we write as the exponential of a time-integrated ``Lagrangian'',
\begin{align}\label{large deviation prob z}
   &\mathbb{P}\Big[ \{z(t), j(t) \}_{t_1 \leq t \leq t_2} = \gamma  \Big] \propto e^{- N \mathcal{A}\left(\gamma \right)}, \qquad  \text{as $N\uparrow \infty$}  
   \end{align}
 {with path-space ``action'' 
   \begin{align*}
  &\mathcal{A}(\gamma) =  - S(z_\text{i}) + \int_{t_1}^{t_2} \id t \ \cal L\big(j(t); z(t) \big) \nonumber 
\end{align*}
Here, the initial state $z_\text{i}$ is sampled from the equilibrium distribution $\propto \exp \left( S(z)\right)$  (where we take $k_B=1$ from now on). Note that we keep in mind a macroscopic system where the large number $N$ counts the number of components or measures the size of the system, but which {\it a priori} allows different trajectories $\gamma$ to be realized.  Such a structure \eqref{large deviation prob z} is assumed, not derived here.\\

The relation between pre-\texttt{GENERIC} \eqref{flow generic} and the probabilities \eqref{large deviation prob z} is taken from the law of large numbers.  We assume that \eqref{flow generic} specifies the trajectory that is overwhelmingly plausible as $N\uparrow \infty$, which is implemented as follows. The ``Lagrangian'' $\cal L(j;z)\geq 0$ is positive, and assumed to be strictly convex in $j$ for all $z$. Consequently, solving $\cal L (j; z) = 0$ for the current $j$ when in state $z$ minimizes the Lagrangian (and hence the action) and gives the most likely current $j = j_z$ which should produce the macroscopic equation \eqref{flow generic}.
In other words, the path-space probabilities \eqref{large deviation prob z}, besides giving the correct dynamical fluctuations, should yield a Lagrangian for which the ``zero-cost flow''
$\cal L(j = j_z; z) = 0$ is equivalent to the pre-\texttt{GENERIC} dynamics \eqref{flow generic}. \\
The function $\cal L$ is known under different names: Lagrangian, rate function, large deviation function, $L$-function, \textit{etc}, \cite{MPR13, largemarkov}. We call it a Lagrangian, not in the sense of mechanics where it is kinetic minus potential energy, but because the equations of motion (here the pre-\texttt{GENERIC} equations) follow from minimizing the path-space action $\cal A = \int \id t \ \cal L$. \\

In what follows, calculations are called ``on shell'' when the zero-cost flow, $j=j_z$, or $\dot{z}(t) =  Dj_{z(t)}$ is assumed.
That (first-order) evolution is equivalent to  $\cal L (j_z;z) = 0, \, \forall z$ and since the Lagrangian is minimal at $j = j_z$ also implies $\partial_j \cal L (j_z;z) = 0, \ \forall z$. It then follows from  
\begin{equation*}
   \forall z : \quad  0 = \frac{\id}{\id z} \Big( \cal L(j_z;z) \Big) = \partial_z \cal L(j_z;z) + \partial_j \cal L(j_z;z) \ \partial_z j_z = \partial_z \cal L(j_z;z) 
\end{equation*}
that also $\partial_z \cal L(j_z;z) = 0 \ \forall z$ and hence the (second-order) Euler-Lagrange equation\footnote{We assume that there are no constraints present in the dynamics except for the `kinematic' relation $\dot z = D j$. See also the example in Section \ref{section nonlinear friction first time} and the corresponding discussion.} is also satisfied, \cite{MPR13},
\begin{align}\label{euler}
    \partial_z \cal L (j_z(t); z(t)) - \frac{\id}{\id t} \Big( \partial_j \cal L (j_z(t); z(t)) \Big) = 0
\end{align}

Note that we did not derive \eqref{euler} from a minimisation of the action, but rather showed that the zero-cost flow $\dot{z}(t) = Dj_{z(t)}$ forms a subclass of the more general Euler-Lagrange equations \eqref{euler}, adding yet another reason for calling $\cal L$ the Lagrangian. \\

 For completeness, we mention here also that the minimization or variational principle for determining the typical and correct evolution can also be obtained from entropy and dissipation principles, at least for quadratic Lagrangians (Gaussian-type processes) when there is no Hamiltonian flow in \eqref{genz} or \eqref{flow generic} (and (pre-)\texttt{GENERIC} reduces to gradient flow). That is done via the Onsager dissipation functions or the Rayleighian and it amounts to a form of minimum or maximum entropy production principle; see \textit{e.g.} \cite{Maesring, mepp, Bruers_2007}. As shown in  \cite{Beyengradientflow}, we find that the zero-cost flow by minimizing the Lagrangian is the same as the equations that follow from extremizing the Rayleighian for Gaussian-type processes. In other words, the Lagrangian in our large deviation setup is similar, but not equivalent, to the  Rayleighian in Onsager's principle, and both lead to the same equations of motion (when there is no reversible part; only gradient flow).\\
However, we want to emphasize that the large-deviation Lagrangian can also be applied to \textit{e.g.} discrete Markov jump processes \cite{lagrangianmarkov} and nonquadratic Lagrangians \cite{Beyengradientflow, genericmaes}, as well as dynamics with a reversible component, which do not (typically) fall under Onsager's principle. Moreover, the path-space action is not only used to derive the equations of motion but also allows for \textit{e.g.} estimating fluctuations.  \\

Lastly, due to the constraint $\dot{z}(t) = D j(t)$, the current $\bar{j}(t)$ associated with the quasistatic trajectories $\bar{z}$ in \eqref{zbar} follows a similar expansion
\begin{align}
    \bar{j}(t) &= j_{\lambda(\ve t)}^* + \ve \ \cal J(\ve t) + O(\ve^2), \qquad j_{\lambda(\ve t)}^* = 0, \qquad D \cal J(\ve t) = \partial_\lambda z^*_{\lambda(\ve t)} \dot{\lambda}(\ve t)
    \label{jbar}
\end{align}
to be used below in the proof of the Noether theorem.
\subsection{Hamiltonian formalism}
The Hamiltonian $\cal H$ is obtained from the Lagrangian $\cal L$ through a Legendre transformation
\begin{equation}\label{hmd}
    \cal H(f;z) = \sup_j \Big\{j \cdot f - \cal L(j;z) \Big\}, \qquad   \cal L(j;z) = \sup_f \Big\{j \cdot f - \cal H(f;z) \Big\}
\end{equation}
where $f$ is dual to the current $j$ and represents a thermodynamic force. As in Hamiltonian mechanics, the thermodynamic force $f$ and the macroscopic variable $z$ are independent variables. \\  Based on its connection to the Lagrangian $\cal L$, we immediately find that $\cal H(f;z)$ is \textit{ strictly} convex in its first argument, and
\begin{align}
    & 
    \cal H(0;z) = \sup_j \Big\{ - \cal L(j;z) \Big\} = -\inf_j \Big\{ \cal L(j;z) \Big\} = 0 \label{h(0,z)} \\
    &
    \sup_f \Big\{j_z \cdot f - \cal H(f;z) \Big\} = \cal L(j_z;z) =  0 \iff  \dot{z}(t) = D j_{z(t)} \label{sup L = 0}
\end{align}
such that the zero-cost flow is equivalent to evaluating the supremum in \eqref{sup L = 0}. From \eqref{h(0,z)}, we see that the supremum is reached at $f = 0$ and from uniqueness of the Legendre transform (since $\cal H$ is smooth and strictly convex), this must be the only solution. Moreover, by extremizing \eqref{sup L = 0}, $j_z$ also solves $j = \partial_f \cal H(f;z)$ such that the zero-cost flow becomes
\begin{equation}\label{zero-cost flow}
     \dot{z} = D j_z \iff \cal L(j_z; z) = 0 \iff j_z = \partial_f \cal H(f;z) \text{ and } f = 0
\end{equation}
Note that $j_z = \partial_f \cal H(f;z)$ is the analogue of the first Hamilton equation with $f = 0$ being a special case of the second Hamilton equation $\dot{f} = - \partial_z \cal H(f;z)$ as
$\partial_z \cal H(0;z) = 0$; see \eqref{h(0,z)}. Thus, similar to how $\cal L(j_z; z) = 0$ also implies the Euler-Lagrange equations, the relations $j_z = \partial_f \cal H(f;z) \text{ and } f = 0$ solve Hamilton equations. \\

Lastly, analogous to \eqref{zbar}, we define the quasistatic trajectories of the force $\bar{f}(t)$ through the expansion
\begin{align}
    \bar{f}(t) &= f_{\lambda(\ve t)}^* + \ve \ F(\ve t) + O(\ve^2), \qquad f_{\lambda(\ve t)}^* = 0 \label{fbar}
\end{align}
where $F(\ve t)$ is left unspecified (similar to $\cal Z(\ve t)$). 

\section{Structure of the Lagrangian and Hamiltonian}\label{section on lagrangian and ham}
In this section, we introduce more structure of the Lagrangian/Hamiltonian formalism, important in the derivation of Section \ref{first time entropy as noether section}.

\subsection{Detailed Balance}\label{detb}
Detailed balance/reversibility is an expression of time-reversal invariance that requires the probability in \eqref{large deviation prob z} of any trajectory $\gamma = (z,j)$ to be equal to that of its time-reversal\footnote{We assume $z$ is even under time-reversal.} $(z, - j)$.  That is achieved by requiring that there is a flow $J^H_z$ so that for all trajectories, 
\begin{align*}
    \int_{t_1}^{t_2} &\id t \ \Big[\cal L\left(J^H_{z(t) }- j(t); z(t) \right) - \cal L\left(J^H_{z(t) } + j(t); z(t) \right) \Big] \\
    & = S(z_{t_2})- S(z_{t_1}) = \int_{t_1}^{t_2} \id t \frac{\id}{\id t} \left( S(z(t)) \right) 
\end{align*}
We refer to \cite{genericmaes, Kraaij_2020} for further discussion. Demanding time-reversal invariance for all times $t_1, t_2$, detailed balance can be written as
\begin{equation}\label{dbe l}
    \cal L(J^H_{z(t) }-j(t);z(t)) - \cal L(J^H_{z(t) }+j(t);z(t)) = \frac{\id}{\id t} \left(S(z(t)) \right) =  j(t) \cdot D^{\dagger} \id S(z(t))
\end{equation}
remembering that always $\dot{z}(t) = D j(t)$. In the Hamiltonian formalism, using \eqref{hmd}, the detailed balance condition reads
\begin{equation}\label{dbe h}
    \cal H(-f(t) - D^{\dagger} \id S(z(t))/2;z(t)) - \cal H(f(t) - D^{\dagger} \id S(z(t))/2; z(t)) = -2 f(t) \cdot J^H_{z(t) }
\end{equation}
The conditions \eqref{dbe l}--\eqref{dbe h} play an important role in proving that entropy is a Noether charge, as they give $\cal L, \cal H$ a structure which we discuss next. This is to be expected since reversibility is one of the conditions for the entropy to remain invariant.

\subsection{Canonical structure}\label{cans}
For the path-space action to generate the dynamics \eqref{flow generic} as zero-cost flow, we need to highlight some extra structure.  It can be shown, \cite{genericmaes, Kraaij_2020}, that the time-symmetric contribution to the Lagrangian can be written as the sum of a Legendre-conjugate pair,
\begin{equation}\label{low}
\frac{\cal L(J^H_{z } + j;z) + \cal L(J^H_{z }-j;z)}{2} = \psi(j;z) +\psi^\star(D^\dagger \id S/2;z)
\end{equation}
where \begin{equation}\label{pst}
\psi^\star(f;z) = \sup_j\left[j\cdot f - \psi(j;z)\right]
\end{equation} 
is the Legendre transform of
\begin{eqnarray}
\psi(j;z) &=& \frac{1}{2}\bigl[\cal L(J^H_{z }-j;z) + \cal L(J^H_{z }+j;z)\bigr]-\cal L(J^H_{z };z) \nonumber\\
 &=& \frac{1}{2}\,j\cdot D^\dagger \id S(z) + \cal L(J^H_{z }+j;z) -\cal L(J^H_{z };z)  \label{Lpsi}
\end{eqnarray}
The second line follows from detailed balance \eqref{dbe l} and implies that $\psi(j;z)$ is convex in $j$. Since $\psi$ is also symmetric in $\pm j$ and vanishes at $j=0$, it follows that $\psi$ is positive, $\psi(j;z)\geq 0$ and minimial at $j = 0$, \textit{i.e.}
\begin{align}\label{partial_j psi = 0}
\forall z: \ \partial_j \psi(0; z) = 0
\end{align}
From \eqref{Lpsi},
\begin{equation}\label{lo} \cal L(J^H_{z }+j;z) = -\frac{1}{2}\,j\cdot D^\dagger \id S(z) + \psi(j;z) +\psi^\star(D^\dagger \id S/2;z)
\end{equation}
which is \eqref{low}.\\
The point is now that the convex $\psi^{\star}$ in \eqref{pst} is the function that appears in \eqref{current flow generic}. 
Indeed, replacing $\psi$ in \eqref{lo} via \eqref{Lpsi}, we see that
\begin{equation}\label{Hpsistar1}
\psi^\star(f;z) = \cal H(f-D^\dagger \id S/2;z) - J^H_{z } \cdot f  + \cal L(J^H_{z };z)
\end{equation}
which is symmetric in $\pm f$, positive and vanishes only at $f=0$ due to the detailed balance conditions \eqref{dbe l}--\eqref{dbe h}. Following the Hamilton equation in \eqref{zero-cost flow}, the zero-cost flow satisfies
\begin{equation}\label{pH2}
   j_z =\partial_f \cal H(0;z)= J^H_z + \partial_f \psi^\star( D^\dagger \id S/2;z) = J_z^H + J_z^S
\end{equation}
As a consequence, the typical path $\dot{z}(t) = D j_{z(t)} =  D (J_{z(t)}^H + J_{z(t)}^S)$ indeed has the pre-{\tt GENERIC} structure \eqref{flow generic}. 

\section{Entropy as Noether charge}\label{first time entropy as noether section}
The structures of the preceding sections can be generalized to include an external protocol $\lambda^{(\varepsilon)}(t) = \lambda(\varepsilon t)$ and leads to a time-dependent Lagrangian $\cal L_{\lambda(\ve t)}$ and Hamiltonian $\cal H_{\lambda(\ve t)}$ with dynamics \eqref{quasistatic gradient flow eq},
\begin{align}
&\dot{z}^{(\ve)}(t) 
  = D j_{z^{(\ve)}(t)} = D J^H_{z^{(\ve)}(t)} 
     + D \, \partial_f \psi^\star\left(
         \frac{1}{2} D^\dagger \, \id S_{\lambda(\varepsilon t)}(z^{(\ve)}(t));
         z^{(\ve)}(t)
     \right) \nonumber \\
  &\iff \mathcal{L}_{\lambda(\varepsilon t)}(j_{z^{(\ve)}(t)} ;z^{(\ve)}(t)) = 0 \nonumber \\
  &\iff j_{z^{(\ve)}(t)} 
      = \partial_f \mathcal{H}_{\lambda(\varepsilon t)}(f = 0; z^{(\ve)}(t)) \nonumber
\end{align}
and fixed initial condition $z^{(\varepsilon)}(t_1) = z^{*}_{\lambda_\text{ini}}$ corresponding to the equilibrium solution at parameter value $\lambda(\tau_1) =\lambda_\text{ini}$.  Note also that $j^{(\ve)}(t) = j_{z^{(\ve)}(t)}$ and $f^{(\ve)}(t) = 0$ for the zero-cost flow. \\ 

We are ready to prove the central result of this paper, namely that the adiabatic invariance of the entropy is a consequence of Noether's theorem for the path-space action $\cal A$, thereby generalizing \cite{Beyengradientflow} to  (pre-){\tt GENERIC}. For the basic strategy, we remind the reader of the end note of Section \ref{qas} saying Noether's theorem comes in two steps: 1) proving the existence of a (quasi)symmetry for a general class of trajectories and 2) imposing the equations of motion.

\subsection{Step 1: quasisymmetry for quasistatic trajectories}
We recall the class of quasistatic trajectories \eqref{zbar}, \eqref{jbar} and \eqref{fbar} in the Lagrangian and Hamiltonian setup:
\begin{align}
\text{(Lagrangian): } &   \bar{z}(t) = z_{\lambda(\ve t)}^* + \ve \  \cal Z(\ve t) + O(\ve^2) , \label{off shell expansion lagrangian} \\
&  \bar{j}(t) =0 + \ve \cal J(\ve t) + O(\ve^2) \nonumber \\
\text{(Hamiltonian): }&  \bar{z}(t) = z_{\lambda(\ve t)}^* + \ve \  \cal Z(\ve t) + O(\ve^2) \label{off shell expansion hamiltonian}\\
& \bar{f}(t) = 0 + \ve \ F (\ve t) + O(\ve^2)  \nonumber 
\end{align}
and we assume that the terms
\begin{align}
\text{(Lagrangian): } &  \partial_j^2 \psi(0, z^*_{\lambda(\ve t)}) \left(  \cal J(\ve t) +  \id J_{z^*_{\lambda(\ve t)}}^H \ \cal Z(\ve t) \right), \qquad \partial_j \partial_z \psi(0; z^*_{\lambda(\ve t)}) \cdot \cal Z(\ve t)  \nonumber  \\
&  \partial^2_{\lambda} S_{\lambda(\ve t)}(z^*_{\lambda(\ve t)}) \cdot  \cal Z(\ve t) \label{bounded conditions 2} \\
\text{(Hamiltonian): }&\partial_z \partial_f \cal H_{\lambda(\varepsilon t)}(0;z^*_{\lambda(\varepsilon t)}) \cdot \cal Z(\ve t),\qquad \partial_f^2 \cal H_{\lambda(\ve t)} (0; z_{\lambda(\ve t)}^{*}) \ F(\ve t) \label{bounded conditions ham}\\
& \partial^2_{\lambda} S_{\lambda(\ve t)}(z^*_{\lambda(\ve t)}) \cdot  \cal Z(\ve t), \qquad \id^2 S_{\lambda(\varepsilon t)}(z^*_{\lambda(\varepsilon t)}) \cdot \cal Z(\ve t)  \nonumber
\end{align}
remain bounded $\ \forall t\in [t_1 = \tau_1/\ve,t_2 = \tau_2/\ve]$ as $\ve \downarrow  0$.\\

Then, for the above subclass of quasistatic trajectories, the continuous symmetry (with $\eta$ infinitesimal)
    \begin{align}
     \text{(Lagrangian) : } t \to t' = t \qquad   &z(t) \to z'(t) = z(t)\nonumber  \\
     & j(t) \to j'(t) = j(t) - 2 \eta \ j(t) \label{shift dl} \\
     \text{(Hamiltonian): } t \to t' = t \qquad   &z(t) \to z'(t) = z(t) \nonumber  \\  
     &f(t) \to f'(t) = f(t) + \eta \ D^{\dagger} \id S_{\lambda(\varepsilon t)}(z(t)) \label{shift df}
    \end{align}
    leaves the action invariant in the quasistatic limit $\varepsilon  \downarrow  0$ in the sense of a quasisymmetry \cite{quasisymmetry}
    \begin{equation}\label{quasisymmetry action}
       \lim_{\ve \downarrow 0} \delta \mathcal{A} = \eta \int_{\tau_1}^{\tau_2} \id \tau \frac{\id }{\id \tau}\left( S_{\lambda(\tau)}(z^*_{\lambda(\tau)}) \right), \qquad \tau = \varepsilon t
    \end{equation}
   with $S_{\lambda}$ from \eqref{current flow generic}. The transformations \eqref{shift dl}--\eqref{shift df} are the same as in \cite{Beyengradientflow}. \\
   The result \eqref{quasisymmetry action} can also be rewritten in the geometric form 
   \begin{align}\label{geometric}
    \lim_{\ve \downarrow 0} \delta \mathcal{A} = \eta \int_{\lambda_{\text{i}}}^{\lambda_{\text{f}}} \id \lambda \ \frac{\id }{\id \lambda}\left( S_{\lambda}(z^*_\lambda) \right)
\end{align}
indicating that, in the quasistatic limit, the change in the action is purely geometrical, invariant under a time reparametrisation $t \to \gamma(t) $ for a smooth function $\gamma(t)$ with $\gamma'(t) > 0$. The intuition here is that the dynamics get so slow as $\ve \downarrow 0$ that the system becomes invariant under time-rescalings.

\subsubsection{Hamiltonian version}\label{proofH}
To prove the assertion \eqref{quasisymmetry action}, we start with the path-space action $\cal A$ in terms of the Hamiltonian,
\begin{equation}\label{action ham}
    \mathcal{A} = - S_{\lambda_{\text{i}}}(z_{t_1}) + \int_{t_1}^{t_2} \id t \ \big[f(t) \cdot j(t) - \cal H_{\lambda(\varepsilon t)}(f(t);z(t)) \big]
\end{equation}
where the thermodynamic force $f$ and the macroscopic state $z$ are independent variables, implying also that $j(t)$, a function of the $z-$trajectory through $\dot z(t) = Dj(t)$, is not changing under the $f$-shift. 
Under the \textit{continuous} symmetry \eqref{shift df}, the action changes as
\begin{align}
    \delta \mathcal{A}  
    &= \eta \int_{t_1}^{t_2} \id t  \Big[ D j(t) \cdot \id S_{\lambda(\varepsilon t)}(z(t)) - \partial_f \cal H_{\lambda(\varepsilon t)}(f(t);z(t)) \cdot D^{\dagger} \id S_{\lambda(\varepsilon t)}(z(t)) \Big] \nonumber \\
    &= \eta \int_{t_1}^{t_2} \id t  \Bigg[\frac{\id}{\id t}\left( S_{\lambda(\varepsilon t)}(z(t)) \right) - \ve \ \dot{\lambda}(\ve t) \cdot \partial_\lambda S_{\lambda(\ve t)}(z(t)) \label{change in action tussenstap}  \\
    & \hspace{5 cm}- D \partial_f \cal H_{\lambda(\varepsilon t)}(f(t);z(t)) \cdot \id S_{\lambda(\varepsilon t)}(z(t)) \Bigg] \nonumber 
\end{align}
Note that $\delta \mathcal{A}$ does not form a total derivative $\int \id t \ \id \Phi/\id t$ for general $(z(t), f(t))$. Focusing instead on the subclass of trajectories \eqref{off shell expansion hamiltonian} only, it follows that
\begin{align*}
&\ve \ \dot{\lambda}(\ve t) \cdot \partial_\lambda S_{\lambda(\varepsilon t)} \left(\bar{z}(t) \right) \\
&= \ve \ \dot{\lambda}(\ve t) \cdot \partial_\lambda S_{\lambda(\varepsilon t)}(z^*_{\lambda(\ve t)}) + \ve^2 \dot{\lambda}(\ve t) \cdot \partial^2_\lambda S_{\lambda(\ve t)}(z^*_{\lambda(\ve t)}) \ \cal Z(\ve t)  + O(\ve^3) \\
& = \ve^2 \dot{\lambda}(\ve t) \cdot \partial^2_\lambda S_{\lambda(\ve t)}(z^*_{\lambda(\ve t)}) \ \cal Z(\ve t)  + O(\ve^3)
\end{align*}
and \\
\begin{align*}
    &D\partial_f \cal H_{\lambda(\varepsilon t)}\left(\bar{f}(t);\bar{z}(t) \right)   \\
    &\qquad = Dj_{z^*_{\lambda(\varepsilon t)}} +  \varepsilon D\left(\partial_f^2 \cal H_{\lambda(\ve t)} (0; z_{\lambda(\ve t)}^{*}) \ F(\ve t) +\partial_z \partial_f \cal H_{\lambda(\varepsilon t)}(0;z^*_{\lambda(\varepsilon t)}) \  \cal Z(\ve t) \right) +  O(\varepsilon^2)  \\
     & \qquad = \varepsilon D\left(\partial_f^2 \cal H_{\lambda(\ve t)} (0; z_{\lambda(\ve t)}^{*}) \ F(\ve t) +\partial_z \partial_f \cal H_{\lambda(\varepsilon t)}(0;z^*_{\lambda(\varepsilon t)}) \  \cal Z(\ve t) \right) +  O(\varepsilon^2)  \\
 &\id S_{\lambda(\varepsilon t)}\left(\bar{z}(t) \right) \\
 & \qquad = \id S_{\lambda(\varepsilon t)}(z^*_{\lambda(\varepsilon t)} ) + \varepsilon  \left( \id^2 S_{\lambda(\varepsilon t)}(z^*_{\lambda(\varepsilon t)}) \ \cal Z(\ve t) \right) + O(\varepsilon^2) \\
 & \qquad =\varepsilon  \left( \id^2 S_{\lambda(\varepsilon t)}(z^*_{\lambda(\varepsilon t)}) \ \cal Z(\ve t) \right) + O(\varepsilon^2) 
\end{align*}
These results follow from the equilibrium conditions \eqref{fixed point z dS}, $j_{z_\lambda^*} = 0$ \eqref{j^* = 0} and the relation $j_z = \partial_f \cal H(0; z)$ \eqref{zero-cost flow}. \\

The change in the action \eqref{change in action tussenstap} then becomes
\begin{align}\label{delta A ds/dt}
     \delta \mathcal{A} &= \eta \int_{t_1}^{t_2} \id t  \left[ \frac{\id}{\id t}\left( S_{\lambda(\varepsilon t)}(\bar{z}(t)) \right) + O(\varepsilon^2) \right] \\
     & = \eta \left( S_{\lambda(\ve t_2)}(\bar{z}(t_2) - S_{\lambda(\ve t_1)}(\bar{z}(t_1)) \right) + \eta \ O(\ve)  \nonumber
\end{align}
where the $O(\ve^2)$ terms inside the integral remain bounded over the entire trajectory due to \eqref{bounded conditions ham} and one should keep in mind that $t_1 = \tau_1/\ve, t_2 = \tau_2/\ve$. Expanding the entropy terms further in $\ve$ using \eqref{zbar} yields
\begin{align*}
     \delta \mathcal{A} & = \eta \left(S_{\lambda(\ve t_2)}(z^*_{\lambda(\ve t_2)}) - S_{\lambda(\ve t_1)}(z^*_{\lambda(\ve t_1)}) \right) \\
     & \qquad + \eta \ \ve \left( \id S_{\lambda(\ve t_2)}(z^*_{\lambda(\ve t_2)}) \  \cal Z(\ve t_2) - \id S_{\lambda(\ve t_1)}(z^*_{\lambda(\ve t_1)}) \  \cal Z(\ve t_1) \right) + \eta \ O(\ve)  \\
     & = \eta \left(S_{\lambda(\ve t_2)}(z^*_{\lambda(\ve t_2)}) - S_{\lambda(\ve t_1)}(z^*_{\lambda(\ve t_1)}) \right) + \eta \ O(\ve) \\
     & = \eta \int_{\tau_1}^{\tau_2} \id \tau  \frac{\id}{\id \tau}\left( S_{\lambda(\tau)}(z^*_{\lambda(\tau)}) \right) + \eta \ O(\ve)
\end{align*}
where we used $\id S_{\lambda(\ve t)}(z^*_{\lambda(\ve t)}) = 0$ in the middle. Taking the quasistatic limit $\varepsilon \to 0$, this reduces to 
\begin{align*}
    \lim_{\ve \downarrow 0} \delta \mathcal{A} =  \eta \int_{\tau_1}^{\tau_2} \id \tau  \frac{\id }{\id \tau} \left( S_{\lambda(\tau)}(z^*_{\lambda(\tau)}) \right)
\end{align*}
which is \eqref{quasisymmetry action}. 

\subsubsection{Lagrangian version}\label{proofL}
In the Lagrangian setup, our starting point is the action
\begin{equation}\label{action l}
    \mathcal{A} = - S_{\lambda_{\text{i}}}(z_{t_1}) +  \int_{t_1}^{t_2} \id t \ \cal L_{\lambda(\varepsilon t)}(j(t); z(t)) 
\end{equation} 
Under the shift \eqref{shift dl}, the action changes as
\begin{align}
    \delta \mathcal{A} &=  -2 \eta\int_{t_1}^{t_2} \id t \ j(t) \cdot \frac{\partial \cal L_{\lambda(\ve t)}}{\partial j}(j(t); z(t)) \\
    &= \eta \int_{t_1}^{t_2} \id t \Big[j(t) \cdot D^{\dagger} \id S_{\lambda(\ve t)}(z(t)) -2 j(t) \cdot \partial_j \psi\left(j(t) - J_{z(t)}^H ;z(t)\right) \Big] \nonumber \\
    &= \eta \int_{t_1}^{t_2} \id t \Bigg[ \frac{\id }{\id t} \left( S_{\lambda(\ve t)}(z(t)) \right) - \ve \ \dot{\lambda}(\ve t) \cdot \partial_\lambda S_{\lambda(\ve t)}(z(t)) \label{change in action lagrangian}  \\
    & \hspace{5 cm}-2 j(t) \cdot \partial_j \psi\left(j(t)  - J^H_{z(t)} ;z(t)\right)\Bigg ] \nonumber
\end{align}
where we used \eqref{orthogonality condition J}-\eqref{lo} in the middle line and \eqref{zdot ds} in the last. This $\delta \mathcal{A}$ does not form a total derivative $\int \id t \ \id \Phi/\id t$ for general $(z(t), j(t))$. Focusing instead on the subclass of trajectories \eqref{off shell expansion lagrangian} only, we find, as before, 
\begin{align*}
&\ve \ \dot{\lambda}(\ve t) \cdot \partial_\lambda S_{\lambda(\varepsilon t)} \left(\bar{z}(t) \right) 
=  \ve^2 \dot{\lambda}(\ve t) \cdot \partial^2_\lambda S_{\lambda(\ve t)}(z^*_{\lambda(\ve t)}) \ \cal Z(\ve t) + O(\ve^3) \\ 
&J^H_{\bar{z}(t)}= J^H_{z_{\lambda(\ve t)}^*} + \ve \ \id J_{z^*_{\lambda(\ve t)}}^H \ \cal Z(\ve t) + O(\ve^2) = \ve \ \id J_{z^*_{\lambda(\ve t)}}^H \ \cal Z(\ve t) + O(\ve^2)
\end{align*}
and
\begin{align*}
   &\bar{j}(t) \cdot \partial_j \psi \left(\bar{j}(t)-J^H_{\bar{z}(t)},\bar{z}(t) \right) =  \ve \ \cal J(\ve t) \cdot  \partial_j \psi(0, z_{\lambda(\ve t)}^*) \\
   & \qquad + \ve^2 \ \cal J(\ve t) \cdot \Bigg[\partial_j \partial_z \psi(0, z^*_{\lambda(\ve t)}) \ \cal Z(\ve t)+ \partial_j^2 \psi(0, z^*_{\lambda(\ve t)}) \left(  \cal J(\ve t) +  \id J_{z^*_{\lambda(\ve t)}}^H \ \cal Z(\ve t) \right) \Bigg] + O(\ve^3) \\
    & \qquad = \ve^2 \ \cal J(\ve t) \cdot \Bigg[\partial_j^2 \psi(0, z^*_{\lambda(\ve t)}) \left(  \cal J(\ve t) +  \id J_{z^*_{\lambda(\ve t)}}^H \ \cal Z(\ve t) \right) +\partial_j \partial_z \psi(0, z^*_{\lambda(\ve t)}) \ \cal Z(\ve t) \Bigg] + O(\ve^3)
\end{align*}
where we have used the equilibrium conditions \eqref{fixed point z dS} and 
$ \partial_j \psi(0;z)  = 0$ \eqref{partial_j psi = 0}. Moreover, due to \eqref{bounded conditions 2}, the $O(\ve^2)$ terms remain bounded over the entire trajectory.  \\

The change in the action \eqref{change in action lagrangian} then becomes
\begin{align}\label{lagrangian delta a dsdt}
     \delta \mathcal{A} &= \eta \int_{t_1}^{t_2} \id t  \left[ \frac{\id }{\id t} \left( S_{\lambda(\varepsilon t)}(\bar{z}(t)) \right) + O(\varepsilon^2) \right] \\
     &=  \eta \left( S_{\lambda(\ve t_2)}(\bar{z}(t_2)) - S_{\lambda(\ve t_1)}(\bar{z}(t_1)) \right) + \eta \ O(\ve)  \nonumber 
\end{align}
since $t_1 = \tau_1/\ve, t_2 = \tau_2/\ve$, which is the same as \eqref{delta A ds/dt} in the Hamiltonian case. Hence, following the same steps, we come to the conclusion
\begin{align*}
    \lim_{\ve \downarrow 0} \delta \mathcal{A} =  \eta \int_{\tau_1}^{\tau_2} \id \tau  \frac{\id}{\id \tau}\left( S_{\lambda(\tau)}(z^*_{\lambda(\tau)}) \right)
\end{align*}
which is \eqref{quasisymmetry action}. 

\subsection{Step 2: quasistatic zero-cost flow}
For the final step in the derivation of the Noether theorem, 
we insert the actual trajectory. {\it i.e.,} evaluating the quasisymmetry for the quasistatic trajectories on shell.

\subsubsection{Hamiltonian version}
Under a general coordinate transform $f \to f + \eta \  \delta f, z \to z $, the action \eqref{action ham} changes as
\begin{equation}\label{df shift}
    \delta \mathcal{A} = \eta \int_{t_1}^{t_2} \id t \  \delta f(t) \left(\dot{z}(t) - \partial_f \cal H(f(t); z(t)) \right) 
\end{equation}
Since \eqref{delta A ds/dt} is equal to  \eqref{df shift}  when $(f(t), z(t)) = (\bar{f}(t), \bar{z}(t))$ for all $t_1<t_2$, the integrands are equal
\begin{equation*}
 \frac{\id }{\id t} \left( S_{\lambda(\ve t)}(\bar{z}(t)) \right) + O(\ve^2) =\delta f(t) \left(\dot{\bar{z}} - \partial_f \cal H \left(\bar{f}(t); \bar{z}(t)\right) \right)
\end{equation*}
which is valid for the entire class \eqref{zbar}. Finally, \textit{on shell}, $ \bar{f}(t) \to  f^{(\ve)}(t) = 0, \  \bar{z}(t) \to z^{(\ve)}(t)$ where $\dot{z}^{(\ve)}(t) = \partial_f \cal H(0 ; z^{(\ve)}(t))$, such that the entropy is conserved in the quasistatic limit $\ve \downarrow  0$ for the physical solution to pre-\texttt{GENERIC}:
\begin{equation*}
S_{\lambda_{\text{f}}}\left(z_{\lambda_\text{f}}^* \right) = S_{\lambda_{\text{i}}}\left(z_{\lambda_\text{i}}^* \right)
\end{equation*}
In summary, we have shown that the entropy function $S_\lambda$ is a constant of motion for the on shell quasistatic reversible trajectories $(f^{(\ve)}(t), z^{(\ve)}(t))$ under the shift \eqref{shift df} in the quasistatic limit $\ve \downarrow 0$.

\subsubsection{Lagrangian version}
Under the transform $j(t) \to j(t) + \eta  \ \delta j(t), z(t) \to z(t) $, the action \eqref{action l} changes as
\begin{equation}\label{general change lagrangian dj shift}
    \delta \mathcal{A} =  \eta \int_{t_1}^{t_2} \id t \  \delta j(t) \cdot \frac{\partial \cal L}{\partial j} (j(t); z(t)) 
\end{equation}
Since \eqref{lagrangian delta a dsdt} is equal to \eqref{general change lagrangian dj shift} when $(j(t), z(t)) = (\bar{j}(t), \bar{z}(t))$ for all $t_1<t_2$, the integrands are equal
\begin{equation*}
 \frac{\id}{\id t} \left( S_{\lambda(\ve t)}(\bar{z}(t)) \right) + O(\ve^2) =\delta j(t) \cdot \frac{\partial \cal L}{\partial j} (\bar{j}(t); \bar{z}(t))
\end{equation*}
which is valid for the entire class \eqref{zbar}. Finally,  \textit{on shell}, $ \bar{j}(t) \to  j^{(\ve)}(t) = j_{ z^{(\ve)}(t)}, \  \bar{z}(t) \to z^{(\ve)}(t)$ where $\frac{\partial \cal L}{\partial j}(j_{z^{(\ve)}(t)}; z^{(\ve)}(t)) =0$, such that the entropy is conserved in the quasistatic limit $\ve \downarrow  0$ for the physical solution to pre-\texttt{GENERIC}:
\begin{equation*}
   S_{\lambda_{\text{f}}}\left(z_{\lambda_\text{f}}^* \right) = S_{\lambda_{\text{i}}}\left(z_{\lambda_\text{i}}^* \right)
\end{equation*}

In summary, we have shown that the entropy function $S_\lambda$ is a constant of motion for the shift \eqref{shift dl} in the quasistatic limit for the on shell quasistatic reversible trajectories $(j^{(\ve)}(t), z^{(\ve)}(t))$. 

\subsection{Additional remarks}
We end this section with some remarks related to the derivation of entropy as a Noether charge. \\

\begin{enumerate}
    \item We repeat here the necessary assumptions/conditions for the invariance of entropy through Noether's theorem: \\
    \begin{itemize}
        \item Reversibility comes from \eqref{dbe l}--\eqref{dbe h} at fixed parameters $\lambda$.  \\
        \item We implement a protocol of the control parameters $\lambda^{(\ve)}(t) = \lambda(\ve t)$ in a quasistatic limit $\ve \downarrow 0$. Moreover, instead of working with a general trajectory $(z(t), j(t))$ (as one usually does in Noether's theorem), we need to focus on the class of quasistatic trajectories \eqref{zbar}. \\
        \item Lastly, at the end of the variation, for the entropy/Noether charge to be conserved, we need to impose the equations of motion (which is the zero-cost flow here) such that $\cal Z(\ve t) \to \zeta(\ve t)$. This selects the physical trajectory from the class \eqref{zbar}. \\
    \end{itemize}
    Regarding adiabicity, as noted in Section \ref{subsection pre generic dynamics}, the $S_\lambda$ in this formalism is only the true entropy for closed, isolated systems, which are adiabatic. Then, implementing the ingredients above yields the equivalent of $\delta Q^{\text{rev}} = 0$ in the first part of Clausius' Heat Theorem. For open systems, the $S_\lambda$ should instead be interpreted as minus the appropriate free energy $\cal F_\lambda$, which does not require the adiabatic condition to be conserved. \\
\item  When the dynamics is purely reversible (\textit{i.e.} no dissipative part in \eqref{flow generic}), it follows from \eqref{dsdt geq 0} that
\begin{align*}
       \frac{\id }{\id t} \left(S_\lambda(z(t)) \right) = 0 \Longrightarrow S_\lambda(z) = \text{constant}, \qquad \id S_\lambda = 0
    \end{align*}
    As $\id S_\lambda = 0$, the symmetry transformation \eqref{shift df} reduces to the identity
    \begin{align*}
        t \to t' = t, \qquad z(t) = z'(t) = z(t), \qquad f(t) \to f'(t) = f(t) 
    \end{align*}
    which has constants as Noether charge, with $S_\lambda$ among them, even without the quasistatic setup. \\
    Alternatively, when starting from a microscopic, mechanical setup, the (phase space) entropy can also be derived as a Noether charge, \cite{Sasa_original, mecos}.
 \\
    \item The Hamiltonian symmetry \eqref{shift df} represents a shift in the thermodynamic force $f(t)$ and is identical to the one in \cite{Beyengradientflow, langevin_noether}. It can also be written in the  form
\begin{align}\label{types}
      t \to t' = t, \qquad   z(t) \to z'(t) &= z(t) - \eta \   D^{\dagger} \{z(t), S_{\lambda(\ve t)}(z(t))\} \\
      f(t) \to f'(t) & = f(t) - \eta  \  D^{\dagger}  \{f(t), S_{\lambda(\ve t)}(z(t))\} \nonumber 
\end{align}
with $\{\cdot , \cdot \}$ the Poisson bracket
\begin{equation*}
    \{F, G\} = \frac{\partial F}{\partial z} \frac{\partial G}{\partial f} - \frac{\partial G}{\partial z} \frac{\partial F}{\partial f}
\end{equation*}
which is the same as \eqref{brackets} for $L = \begin{bmatrix}
    0 & 1 \\
    -1 & 0
\end{bmatrix}$. In other words, the entropy generates its own symmetry in the Hamiltonian formalism, as also explained and applied in \cite{Beyengradientflow, langevin_noether, laghamconnection}. The same symmetry transform appeared in \cite{mecos} for the derivation of the (phase space) entropy as a Noether charge in a mechanical setup, connecting our mesoscopic derivation to that of a microscopic, mechanical framework. \\
\item For the Lagrangian setup, the relevant symmetry \eqref{shift dl} becomes a shift in the current $j$ without changing the macroscopic variable $z$, {\it i.e.}, $\delta z(t) = 0$. As such, the transformations of the generalized `velocities' are not determined by those of the coordinates $z$ and the time $t$. This kind of transformation is typically not considered for Noether's theorem, as it does not lead to a symmetry of the action. However, this type of transformation has appeared in \cite{nonconservativenoether, Vittal1988} when describing Noether's theorem for mechanical systems subject to nonconservative forces. Moreover, due to the duality between the force $f$ and current $j$ from Section \ref{section on lagrangian and ham}, one expects that a shift in the force \eqref{shift df} should correspond to a shift in the current.
\end{enumerate}
   
\section{Example: nonlinear friction}\label{section nonlinear friction first time}

We end the paper with an example from \cite{genericmaes}. Not only does it serve as a concrete illustration for the more formal analysis of the previous sections, but it also adds some extensions and remarks that would remain less visible in the general treatment.  We mention in particular: (1) that the Hamiltonian/reversible flow in (pre-){\tt GENERIC} can more generally also be affected by the quasistatic protocol $\lambda^{(\ve)}(t)$ 
(\textit{e.g.} because the energy function depends on slowly changing parameters $\lambda$), (2) what happens when the dissipative part $J_z^S$ vanishes in pre-{\tt GENERIC} and (3) how to incorporate (mechanical) constraints.

\subsection{The dynamics}
This simple example has operator $D=$ the identity in \eqref{flow generic} and gives the dynamics for a macroscopic particle with mass $m$ and states $z(t) = (p(t),q(t))^T$ moving in the two-dimensional phase space according to
\begin{equation}\label{nonfric}
    \dot{q}(t) = \frac{p(t)}{m}, \qquad \dot{p}(t) = - V'(q(t)) - 2 \varphi \sinh \left(\frac{p(t)}{2mv_o} \right)
\end{equation}
It exhibits nonlinear friction with strength $\varphi>0$, where $v_o>0$ is some reference speed\footnote{ 
Eq. \eqref{nonfric} can be derived as the $N \uparrow \infty$ limit of a particle with mass $m$, randomly colliding and exchanging momentum with $N$ smaller particles with velocity $2 v_0$ and mass $\tilde{m}/N$, \cite{genericmaes}.}, while moving in a strictly convex and confining potential $V$.  The parameter $\varphi$ is useful for following the dependence of the arguments on the dissipative part in the dynamics.\\
Eq. \eqref{nonfric} can be written in the pre-\texttt{GENERIC} form \eqref{flow generic} for the current $j_z = J^H_z + J_z^S$ which 
splits into a Hamiltonian $J_z^H$ and a dissipative $J_z^S$ part
\begin{align*}
      J^H_z & = \left(p/m, - V'(q) \right)^T, \qquad        J_z^S = \left(0,-2 \varphi \sinh  \left(\frac{p}{2mv_o} \right) \right)^T 
      \end{align*}
The dissipative part $J_z^S$ can be written in the form \eqref{current flow generic},
\begin{align}
   &J_s^S =  \nabla_f \psi^{\star}\left(D^{\dagger} \partial_q \left(- \frac{\cal F(q,p)}{2 m v_0^2}\right), D^{\dagger} \partial_p \left(- \frac{\cal F(q,p)}{2 m v_0^2}\right) ;q,p\right) \nonumber \\
      &\psi^{\star}(f_q, f_p; q,p) = \frac{2 \varphi}{m v_0} (\cosh(m v_0 f_p) - 1) \label{psi example early} \\
      &\cal F(p,q) = p^2/2m + V(q) \nonumber
\end{align}
with $\nabla_f = \left(\partial_{f_q}, \partial_{f_p} \right)^T$. We recognize the entropy $S = -\cal F/m v_0^2$ as a dimensionless free energy $\cal F$, which is to be expected as we are dealing with an open system. Along \eqref{nonfric},
\[
\frac{\id}{\id t}{\cal F}(z(t)) = \frac{p(t)}{m}\dot p(t) +  V'(q(t))\ \dot q(t) = -2\varphi\,\frac{p(t)}{m}\,\sinh \left(\frac{p(t)}{2mv_o} \right) \leq 0
\]
indicating that the (mechanical) energy is not conserved, \textit{i.e.} we are outside \texttt{GENERIC} (but inside pre-\texttt{GENERIC}). 
The thermodynamic interpretation is that mechanical energy is converted into heat. Therefore, to have a conserved energy as needed in \texttt{GENERIC} and as explained in \cite{Kraaij_2020}, we would have to add an ‘internal energy’ variable $e$ that captures this heat with enlarged state space.

  The dynamics of the free energy $\cal F$ of the system is equivalent to the behavior of the entropy of the total system, which is the particle plus heat bath. \\

Parameters $\lambda$ can be added to the potential $V$,  and in that sense, the Hamiltonian flow $J_z^H$ is directly affected by $\lambda$ since $V = V_\lambda(q)$.  The macroscopic equilibrium state $(q_\lambda^*, p_\lambda^*)$ satisfies $\partial_q \cal F_\lambda(q_\lambda^*, p_\lambda^*) = \partial_p \cal F_\lambda(q_\lambda^*, p_\lambda^*) = 0, J_{z^*_\lambda}^H = 0$, or $p_\lambda^* = 0, V'_\lambda(q_\lambda^*) = 0$ such that the particle remains at the minimum of the potential $V_\lambda$ with zero momentum. Clearly, one then has $\dot{q}_\lambda^*, \dot{p}_\lambda^* = 0$ in \eqref{nonfriction} and the free energy is minimal $\cal F^*_\lambda = V(q_\lambda^*)$. Moreover, we require the potential to be such that $\partial_\lambda \cal F_\lambda(q_\lambda^*)=  \partial_\lambda V_\lambda(q_\lambda^*) = 0$. \\

Next, we make the control parameter $\lambda \to \lambda(\ve t)$ time-dependent for a small rate $\ve > 0$, with instantaneous equilibria $q^*_{\lambda(\ve t)}, p^*_{\lambda(\ve t)}$ and write the quasistatic trajectories \eqref{zbar} as
\begin{align}\label{qbar pbar}
    \bar{q}(t) = q_{\lambda(\ve t)}^* + \ve \cal Q(\ve t) + O(\ve^2), \qquad  \bar{p}(t) = p_{\lambda(\ve t)}^* + \ve \cal P(\ve t) + O(\ve^2)
\end{align}
On the other hand, the zero-cost flow \eqref{quasistatic gradient flow eq} becomes
\begin{equation}\label{nonlinear friction lambda eom}
     \dot{q}^{(\ve)}(t) = \frac{p^{(\ve)}(t)}{m}, \qquad \dot{p}^{(\ve)}(t) = - V'_{\lambda}(q) \left (q^{(\varepsilon)}(t) \right) - 2 \varphi \sinh \left(\frac{p^{(\ve)}(t)}{2 m v_0} \right)
\end{equation}
When starting from equilibrium $q^{(\ve)}(t_1) = q_{\lambda(\ve t_1)}^*, p^{(\ve)}(t_1) = p_{\lambda(\ve t_1)}^*$, the solution of \eqref{nonlinear friction lambda eom} will slowly deviate from it. Indeed, perturbatively in $\ve$, we have \eqref{expa},
\begin{align}\label{quasistatic trajectories non fric}
    q^{(\varepsilon)}(t) & = q^{*}_{\lambda(\varepsilon t)} + \varepsilon \zeta_q(\ve t) + O(\varepsilon^2), \qquad p^{(\varepsilon)}(t) = p^{*}_{\lambda(\varepsilon t)} + \varepsilon \zeta_p(\ve t) + O(\varepsilon^2)
\end{align}
satisfying
\begin{align*}
   \zeta_q(\ve t) = & -\frac{ \varphi \dot{\lambda}(\tau) \partial_\lambda q^*_{\lambda(\tau)}}{v_0 V''_{\lambda(\ve t)}(q^*_{\lambda(\ve t)})}, \qquad \zeta_p(\ve t) =   m \dot{\lambda}(\tau) \ \partial_{\lambda} q^*_{\lambda(\tau)}
\end{align*}
where we have used that $p^*_\lambda = 0$ and thus also $\partial_\lambda p^*_\lambda = 0$. 
Note that $\zeta_q(\ve t)$ remains bounded only when 
$V''_{\lambda(\tau)}(q^*_{\lambda(\tau)})$ and $v_0$ do not vanish, which means we are not allowed to pass inflection points.    Moreover, when $\varphi=0$, $\zeta_q(\ve t) =0$, while in the overdamped limit $m \downarrow 0, \zeta_p(\ve t) = 0$. \\

\subsection{Lagrangian and Hamiltonian}
The Lagrangian and Hamiltonian for the nonlinear friction dynamics \eqref{nonfric} are derived in \cite{genericmaes}. As in sections \ref{section zero-cost flow} and \ref{section on lagrangian and ham}, we write the equations without the (time-dependent) control parameters $\lambda$. The protocol $\lambda^{(\ve)}(t)$ will be reinstated in the next subsection on Noether's theorem. \\

The Hamiltonian turns out to be\footnote{We have corrected the expression in \cite{genericmaes} by replacing $f_p$ with $- f_p$ inside the $\cosh$ and inserting the correct units.}
\begin{align}
    \cal H(f_q,f_p; q,p) &= \frac{2 \varphi}{m v_0} \cosh \left(- m v_0 f_p + \frac{p}{2 m v_0} \right) - \frac{2 \varphi}{m v_0} \cosh \left(\frac{p}{2 m v_0} \right) \label{ham nonlinear friction} \\
    &  \quad + f_q \frac{p}{m} - f_p V'(q)  \nonumber
\end{align}
with $f_q, f_p$ the conjugate variables to $(q,p)$. The zero-cost flow is \eqref{zero-cost flow}
\begin{align*}
    \dot{q}(t) & = \partial_{f_q} \cal H(0,0; q(t),p(t)) = \frac{p(t)}{m} \\
    \dot{p}(t) & = \partial_{f_p} \cal H(0,0; q(t),p(t)) = - V'(q(t)) - 2 \varphi \sinh \left(\frac{p(t)}{2 m v_0} \right)
\end{align*}
which agrees with \eqref{nonfric}. The full Hamilton equations of motion are
\begin{align}
   \dot{q}(t) & = \partial_{f_q} \cal H =  \frac{p(t)}{m} \\
    \dot{p}(t) &= \partial_{f_p} \cal H = - V'(q(t)) - 2 \varphi \sinh \left(-m v_0 f_{p}(t) + \frac{p(t)}{2 m v_0} \right) \label{nonfriction} \\
    \dot{f}_{q}(t) & = - \partial_q \cal H  =- f_p(t) \  V''(q(t)) \nonumber \\
    \, \dot{f}_{p}(t)  &  = - \partial_p \cal H =  -\frac{\varphi}{(m v_0)^2} \Bigg[ \sinh \left(- m v_0 f_{p(t)} + \frac{p(t)}{2 m v_0} \right) - \sinh \left(\frac{p(t)}{2m v_0} \right) \Bigg] - \frac{f_{q}(t)}{m}  \nonumber
\end{align}
and we regain \eqref{nonfric} when $f_{p(t)},f_{q(t)} = 0$, such that the zero-cost flow is a subclass of the full Hamilton equations, but not {\it vice versa}.  \\

Alternatively, from \cite{genericmaes}, we can immediately write down the Lagrangian\footnote{We corrected the expression in \cite{genericmaes} by adding the last term $ \frac{V'(q)}{2 m v_0^2} \dot{q}$ and inserting the correct units.}
\begin{align}\label{Lagrangiaan friction}
    \mathcal{L}(\dot{q}, \dot{p}; q,p) &= \frac{2 \varphi}{m v_0} \ \Lambda\left( \frac{\dot{p} + V'(q)}{2 \varphi} \right) + \frac{2 \varphi}{m v_0} \cosh \left( \frac{p}{2 m v_0} \right) + \dot{p} \frac{p}{2 m^2 v_0^2} + \frac{V'(q)}{2 m v_0^2} \dot{q} \\
    \Lambda(\gamma) &:=   \gamma \log \left(\gamma + \sqrt{1 + \gamma^2} \right) - \sqrt{1 + \gamma^2} = \gamma \arcsinh(\gamma) - \sqrt{1+\gamma^2} \nonumber
\end{align}
with constraint $\dot{q} = p/m$. Note that the macroscopic variable $z$ now consists of $(q,p)$ with corresponding current $j = (\dot{q}, \dot{p})$. The forces $f_q,f_p$ conjugate to $j$ are not present in the Lagrangian, but were present in the Hamiltonian \eqref{ham nonlinear friction}. \\
 For $\varphi\downarrow 0$, the first term in the Lagrangian diverges logarithmically except when $\dot p = -V'(q)$, which is the equation of motion $\dot{z} = D J_z^H$ (only). It means that the zero-cost flow (of course) better approximates the reversible flow as the dissipative part vanishes. \\ 

To see that the zero-cost flow \eqref{nonfric} follows from $\cal L(\dot{q}, \dot{p}; q,p) = 0$, we note that $\Lambda(-\sinh(x)) = x \sinh(x) - \cosh(x)$. Consequently, $\cal L(\dot{q}, \dot{p}; q,p) = 0$ (only) for $\dot{p} = - V'(q) - 2 \varphi \sinh(p/(2m v_0))$  and  $\dot{q} = p/m$, which is exactly the macroscopic equation \eqref{nonfric}. 
To derive the full Euler-Lagrange equations \eqref{euler} for \eqref{Lagrangiaan friction}, we should take the constraint $\dot{q} = p/m$ into account. The simplest strategy then is to replace $\dot{q}$ with $p/m$ in \eqref{Lagrangiaan friction}, leading to\footnote{More generally, the constraint $f(\dot{q}, p) = \dot{q}(t) - p(t)/m = 0$ is nonholonomic \cite{goldstein2002classical} but linear in the velocities. In classical mechanics, one incorporates them most easily through the d'Alembert-Lagrange principle, which, for nonholonomic constraints, is not derivable from the stationarity of an action $\delta \cal A = 0$ and thus inadequate for our purposes. Differently, in vakonomic mechanics, one adds a Legendre multiplier $\mu(t)$ to the Lagrangian $\cal L \to \tilde{\cal L}(q,p,\dot{q}, \dot{p}, \mu) = \cal L(q, p, \dot{q}, \dot{p}) + \mu f(\dot{q}, p)$ and calculates the corresponding constrained Euler-Lagrange equations from the modified action $\tilde{\cal A} = \int \id t \ \tilde{\cal{L}}$ 
\begin{align*}
    & \frac{\id }{\id t} \left( \frac{\id \cal L}{\id \dot{q}} \right) - \frac{\partial \cal L}{\partial q} = \mu \frac{\partial f}{\partial q} - \frac{\id}{\id t} \left( \mu \frac{\partial f}{\partial \dot{q}} \right) \iff - \dot{\mu}(t) = - \frac{V''(q(t))}{m v_0} \arccosh{\left( \frac{2 \varphi}{\dot{p}(t) + V'(q(t))} \right)} \\
    &\frac{\id }{\id t} \left( \frac{\id \cal L}{\id \dot{p}} \right) - \frac{\partial \cal L}{\partial p} = \mu \frac{\partial f}{\partial p} - \frac{\id}{\id t} \left( \mu \frac{\partial f}{\partial \dot{p}} \right)  \iff - \frac{\mu(t)}{m} = - \frac{\varphi}{m^2 v_0^2} \sinh \left( \frac{p(t)}{2 m v_0} \right) + \frac{\ddot{p}(t) + \dot{q}(t) \  V''(q(t))}{m v_0 \sqrt{4 \varphi^2 + (\dot{p}(t) + V'(q(t)))^2}} \\
    & \frac{\id }{\id t} \left( \frac{\id \cal L}{\id \dot{\mu}} \right) - \frac{\partial \cal L}{\partial \mu} = 0 \iff f(\dot{q}, p) = \dot{q}(t) - p(t)/m  = 0
\end{align*}
One readily checks that \eqref{nonfric} solve these equations for $\mu(t) =- V'(q(t))/(2 m v_0^2)$ (but they don't have to be the unique solution). Plugging this $\mu$ into $\tilde{\cal L}$ yields the same Lagrangian as \eqref{constraint l friction}. See \cite{nonholonomic} for further discussion and comparison between the different strategies.}
\begin{align}
 \mathcal{L}(\dot{p}; q,p) = \frac{2 \varphi}{m v_0} \ \Lambda\left( \frac{\dot{p} + V'(q)}{2 \varphi} \right) &+ \frac{2 \varphi}{m v_0} \cosh \left( \frac{p}{2 m v_0} \right) + \dot{p} \frac{p}{2 m^2 v_0^2} + \frac{V'(q)}{2 m^2 v_0^2} p \label{constraint l friction} \\
     \frac{\id }{\id t} \left( \frac{\id \cal L}{\id \dot{p}} \right) - \frac{\partial \cal L}{\partial p} = 0 \iff 0 &= - \frac{V'(q(t))}{2 m^2 v_0^2} - \frac{\varphi}{m^2 v_0^2} \sinh \left( \frac{p(t)}{2 m v_0} \right) \nonumber \\
     & \quad + \frac{\ddot{p}(t) + \dot{q}(t) V''(q(t))}{m v_0 \sqrt{4 \varphi^2 + (\dot{p}(t) + V'(q(t)))^2}} \nonumber
\end{align}
with $ \dot{q}(t) = p(t)/m $. 
One readily checks that the zero-cost flow \eqref{nonfric} solves these equations (but they don't have to be the unique solution). \\

The Lagrangian and Hamiltonian can be decomposed along equations \eqref{lo} and \eqref{Hpsistar1}, with functions
\begin{equation*}
    \psi(\dot{q}, \dot{p}; q,p) = \frac{2 \varphi}{m v_0} \left(  \Lambda \left( \frac{\dot{p}}{2 \varphi} \right) + 1 \right), \qquad \psi^{\star}(f_q, f_p; q,p) = \frac{2 \varphi}{m v_0} (\cosh(m v_0 f_p) - 1)
\end{equation*}
where $\psi^{\star}$ is the same as in \eqref{psi example early}. 
One easily checks that these functions are convex, positive, symmetric in $\pm j, \pm f$ and vanish at $j = 0, f = 0$ respectively. Lastly, the detailed balance conditions (\ref{dbe l}, \ref{dbe h}) are satisfied
\begin{align*}
      \cal L(J_{z}^H-j;z) - \cal L(J_{z}^H+j;z) = - \dot{p} \frac{p}{m^2 v_0^2} - \frac{V'(q) \dot{q}}{m v_0^2} &= j \cdot D^{\dagger} \id \left(- \frac{\cal F(q,p)}{m v_0^2}\right) \\
      \cal H(-f - D^{\dagger} \id (- \cal F/ m v_0^2)/2;z) - \cal H(f - D^{\dagger} \id (- \cal F/ m v_0^2)/2; z) &= - 2 \frac{p}{m} f_q + 2 f_p V'(q) \\
      & = - 2 f \cdot J_{z}^H
\end{align*}

\subsection{Entropy as a Noether charge}
We demonstrate here that the free energy $\cal F_\lambda$ is a Noether charge under the transformations \eqref{shift dl}--\eqref{shift df}. We give the Lagrangian version.\\

First, we rewrite the Lagrangian \eqref{Lagrangiaan friction} using the free energy $\cal F_{\lambda}$ as
\begin{align*}
   \mathcal{L}_{\lambda(\ve t)}(\dot{q}, \dot{p}; q,p) &= \frac{2 \varphi}{m v_0} \ \Lambda\left( \frac{\dot{p} + V_{\lambda(\ve t)}'(q)}{2 \varphi} \right) + \frac{2 \varphi}{m v_0} \cosh \left( \frac{p}{2 m v_0} \right) \\
   &\quad + \dot{p} \frac{\partial_p \cal F_\lambda(\ve t)}{2 m v_0^2} + \dot{q} \frac{\partial_q \cal F_\lambda(\ve t)}{2 m v_0^2}
\end{align*}
where we inserted the protcol $\lambda(\ve t)$. Following \eqref{shift dl}, the relevant symmetry becomes a shift in the current $j = (\dot{q}, \dot{p}))$, {\it i.e.}, $\dot{q}(t) \to \dot{q}(t) + \eta \,\delta \dot{q}(t), \dot{p}(t) \to \dot{p}(t) + \eta \,\delta \dot{p}(t)$, without changing the macroscopic variables $(p,q)$, \textit{i.e.} $\delta q(t) = \delta p(t) = 0$. As such, to linear order in $\eta$.
\begin{align*}
  \delta \mathcal{A} &= \int_{t_1}^{t_2} \id t \  \delta \cal L_{\lambda(\ve t)} \\
  \delta \cal L_{\lambda(\ve t)} & = \eta \ \frac{\partial \cal L_{\lambda(\ve t)}}{\partial \dot{q}} \delta \dot{q}(t) +  \eta \ \frac{\partial \cal L_{\lambda(\ve t)}}{\partial \dot{p}} \delta \dot{p}(t) \\
     & = \frac{\eta}{2 m v_0^2} \partial_q \cal F_{\lambda(\ve t)} \ \delta \dot{q}(t) + \eta \left( \frac{1}{m v_0} \Lambda'\left( \frac{\dot{p}(t) + V_{\lambda(\ve t)}'(q(t))}{2 \varphi} \right) + \frac{\partial_p \cal F_{\lambda(\ve t)}}{2 mv_0^2} \right) \delta \dot{p}(t) \\
     \Lambda'(\gamma)&= \arcsinh(\gamma)
\end{align*}
Upon choosing $\delta \dot{q} = - 2 \dot{q}, \delta \dot{p} = - 2 \dot{p}$ as in \eqref{shift dl}, this reduces to
\begin{align*}
    \delta \cal L_{\lambda(\ve t)} &= -\eta \ \frac{\partial_q \cal F_{\lambda(\ve t)} \ \dot{q}(t) + \partial_p \cal F_{\lambda(\ve t)} \  \dot{p}(t)}{m v_0^2} - 2 \eta \frac{\dot{p}(t)}{m v_0} \ \Lambda'\left( \frac{\dot{p}(t) + V_{\lambda(\ve t)}'(q(t))}{2 \varphi} \right) \\
    & = \eta \ \frac{\id }{\id t} \left(-\frac{\cal F_{\lambda(\ve t)}}{m v_0^2}\right) + \eta \ \ve \frac{\partial_\lambda \cal F_{\lambda(\ve t)}}{m v_0^2} \dot{\lambda}(\ve t) - 2 \eta \ \frac{\dot{p}(t)}{m v_0} \ \Lambda'\left( \frac{\dot{p}(t) + V_{\lambda(\ve t)}'(q(t))}{2 \varphi} \right)
\end{align*}
Substituting for $(q,p, \dot{q}, \dot{p})$ the quasistatic trajectories \eqref{qbar pbar}, we get
\begin{align}
      \delta \mathcal{A} &= \eta \int_{t_1}^{t_2} \id t \left[ \frac{\id }{\id t} \left(-\frac{\cal F_{\lambda(\ve t)}}{m v_0^2}\right) + O(\ve^2) \right] = \eta \left( -\frac{\cal F_{\lambda_{\text{f}}} - \cal F_{\lambda_{\text{i}}}}{m v_0^2} + O(\ve) \right) \nonumber
\end{align}
and thus
\begin{align*}
     \lim_{\ve \downarrow 0}  \delta \mathcal{A} & = -\eta \ \frac{\cal F_{\lambda_{\text{f}}} - \cal F_{\lambda_{\text{i}}}}{m v_0^2}
\end{align*}
Consequently, the entropy function (here the dimensionless free energy $S_\lambda = -\cal F_\lambda/m v_0^2$) is the Noether charge of a current shift $\delta j = - 2 \eta\ j = - 2 \eta \ (\dot{q}, \dot{p})$ when the variables are following the quasistatic zero-cost flow. \\

\section{Conclusion}
 The first part of Clausius’ Heat Theorem can be linked to Noether’s theorem, identifying thermodynamic entropy as a Noether charge. This connection provides a formal mechanical interpretation of entropy invariance during reversible, adiabatic transformations. We extend this result to macroscopic systems whose evolution is described by the \texttt{GENERIC} framework, unifying Hamiltonian (reversible) and dissipative (irreversible) dynamics in a thermodynamically consistent way. The analysis is carried out in both the Lagrangian and Hamiltonian formalisms, where the energy and entropy functionals also serve as action weights in the path-space formulation of dynamical fluctuation theory. In this setting, the macroscopic evolution equations emerge as zero-cost flows, corresponding to the most probable trajectories. We made the calculations explicit through the example of an inertial probe with nonlinear friction. 

\section*{Statements and Declarations}
AB is supported by the Research Foundation - Flanders (FWO) doctoral fellowship 1152725N.

\bibliographystyle{unsrt}
\bibliography{bib.bib}

\begin{thebibliography}{10}

\bibitem{clausius1865}
R.~Clausius.
\newblock {Ueber verschiedene für die Anwendung bequeme Formen der Hauptgleichungen der mechanischen Wärmetheorie}.
\newblock {\em Annalen der Physik}, 125:353--400, 1865.

\bibitem{ruelle}
D.~Ruelle.
\newblock {Extending the Definition of Entropy to Nonequilibrium Steady States}.
\newblock {\em Proceedings of the National Academy of Sciences of the United States of America}, 100(6):3054--3058, 2003.

\bibitem{jonalasinio2023clausius}
G.~Jona-Lasinio.
\newblock {On Clausius’ approach to entropy and analogies in non-equilibrium}.
\newblock {\em Ensaios Matemáticos}, 38, 2023.

\bibitem{Saito_2011}
K.~Saito and H.~Tasaki.
\newblock {Extended Clausius Relation and Entropy for Nonequilibrium Steady States in Heat Conducting Quantum Systems}.
\newblock {\em Journal of Statistical Physics}, 145(5):1275–1290, 2011.

\bibitem{prigo}
R.J. Donnelly, R.~Herman, and I.~Prigogine.
\newblock {\em {Non-equilibrium Thermodynamics: Variational Techniques and Stability}}.
\newblock University of Chicago Press, 1966.

\bibitem{heatconduction}
T.~S. Komatsu, N.~Nakagawa, S.~Sasa, and H.~Tasaki.
\newblock {Steady-State Thermodynamics for Heat Conduction: Microscopic Derivation}.
\newblock {\em Physical Review Letters}, 100(23):230602, 2008.

\bibitem{Komatsu_2010}
T.~S. Komatsu, N.~Nakagawa, S.~Sasa, and H.~Tasaki.
\newblock {Entropy and Nonlinear Nonequilibrium Thermodynamic Relation for Heat Conducting Steady States}.
\newblock {\em Journal of Statistical Physics}, 142(1):127–153, 2010.

\bibitem{Bertini_2012}
L.~Bertini, D.~Gabrielli, G.~Jona-Lasinio, and C.~Landim.
\newblock {Thermodynamic Transformations of Nonequilibrium States}.
\newblock {\em Journal of Statistical Physics}, 149(5):773–802, 2012.

\bibitem{clausiusstationarystates}
L.~Bertini, D.~Gabrielli, G.~Jona-Lasinio, and C.~Landim.
\newblock {Clausius Inequality and Optimality of Quasistatic Transformations for Nonequilibrium Stationary States}.
\newblock {\em Physical Review Letters}, 110(2):020601, 2013.

\bibitem{Maes_2015}
C.~Maes and K.~Netočný.
\newblock {Revisiting the Glansdorff–Prigogine Criterion for Stability Within Irreversible Thermodynamics}.
\newblock {\em Journal of Statistical Physics}, 159(6):1286–1299, 2015.

\bibitem{Sasa_original}
S.~Sasa and Y.~Yokokura.
\newblock {Thermodynamic Entropy as a Noether Invariant}.
\newblock {\em Physical Review Letters}, 116(14):140601, 2016.

\bibitem{mecos}
A.~Beyen and C.~Maes.
\newblock {The first part of Clausius’ heat theorem in terms of Noether’s theorem}.
\newblock {\em Mathematics and Mechanics of Complex Systems}, 13(1):1–24, 2025.

\bibitem{Noether1918}
E.~Noether.
\newblock {Invariante Variationsprobleme}.
\newblock {\em Nachrichten von der Gesellschaft der Wissenschaften zu Göttingen, Mathematisch-Physikalische Klasse}, 1918:235--257, 1918.

\bibitem{Beyengradientflow}
A.~Beyen and C.~Maes.
\newblock {Entropy as Noether charge for quasistatic gradient flow}.
\newblock {\em Journal of Non-Equilibrium Thermodynamics}, 50(2):241–261, 2025.

\bibitem{langevin_noether}
Y.~Minami and S.~Sasa.
\newblock {Thermodynamic entropy as a Noether invariant in a Langevin equation}.
\newblock {\em Journal of Statistical Mechanics: Theory and Experiment}, 2020(1):013213, 2020.

\bibitem{genericmaes}
R.~Kraaij, A.~Lazarescu, C.~Maes, and M.~Peletier.
\newblock {Deriving GENERIC from a Generalized Fluctuation Symmetry}.
\newblock {\em Journal of Statistical Physics}, 170(3):492–508, 2018.

\bibitem{ottinger2005beyond}
H.C. {\"O}ttinger.
\newblock {\em {Beyond Equilibrium Thermodynamics}}.
\newblock Wiley, 2005.

\bibitem{tanaka2021noetherslearningdynamicsrole}
H.~Tanaka and D.~Kunin.
\newblock {Noether’s Learning Dynamics: Role of Symmetry Breaking in Neural Networks}.
\newblock In {\em {Advances in Neural Information Processing Systems}}, volume~34, pages 25646--25660. Curran Associates, Inc., 2021.

\bibitem{zhao2023symmetriesflatminimaconserved}
B.~Zhao, I.~Ganev, R.~Walters, R.~Yu, and N.~Dehmamy.
\newblock {Symmetries, Flat Minima, and the Conserved Quantities of Gradient Flow}.
\newblock In {\em {The Eleventh International Conference on Learning Representations }}, 2023.

\bibitem{PavelkaKlikaGrmela+2018}
M.~Pavelka, V.~Klika, and M.~Grmela.
\newblock {\em {Multiscale Thermo-Dynamics}}.
\newblock De Gruyter, Berlin, Boston, 2018.

\bibitem{Grmela_2018}
M.~Grmela.
\newblock {GENERIC guide to the multiscale dynamics and thermodynamics}.
\newblock {\em Journal of Physics Communications}, 2(3):032001, 2018.

\bibitem{DZYALOSHINSKII198067}
I.E. Dzyaloshinskii and G.E. Volovick.
\newblock Poisson brackets in condensed matter physics.
\newblock {\em Annalen der Physik}, 125(1):67--97, 1980.

\bibitem{Grmela1984}
M.~Grmela.
\newblock Particle and bracket formulations of kinetic equations.
\newblock {\em Contemporary Mathematics}, 28, 01 1984.

\bibitem{MORRISON1984423}
P.~J. Morrison.
\newblock Bracket formulation for irreversible classical fields.
\newblock {\em Physics Letters A}, 100(8):423--427, 1984.

\bibitem{KAUFMAN1984419}
A.~N. Kaufman.
\newblock {Dissipative Hamiltonian Systems: A Unifying Principle}.
\newblock {\em Physics Letters A}, 100(8):419--422, 1984.

\bibitem{complexfluids1}
M.~Grmela and H.~C. \"Ottinger.
\newblock {Dynamics and thermodynamics of complex fluids. I. Development of a general formalism}.
\newblock {\em Physical Review E}, 56:6620--6632, 1997.

\bibitem{complexfluids2}
H.~C. \"Ottinger and M.~Grmela.
\newblock {Dynamics and thermodynamics of complex fluids. II. Illustrations of a general formalism}.
\newblock {\em Physical Review E}, 56:6633--6655, 1997.

\bibitem{zwanzig}
R.~Zwanzig.
\newblock {Ensemble Method in the Theory of Irreversibility}.
\newblock {\em The Journal of Chemical Physics}, 33(5):1338--1341, 1960.

\bibitem{zwanzig2001nonequilibrium}
R.~Zwanzig.
\newblock {\em {Nonequilibrium Statistical Mechanics}}.
\newblock Oxford University Press, 2001.

\bibitem{bouchard2007morizwanzigequationstimedependentliouvillian}
L.-S. Bouchard.
\newblock {Mori-Zwanzig Equations With Time-Dependent Liouvillian}, 2007.
\newblock arXiv:0709.1358 [physics.chem-ph].

\bibitem{ottingerprojection1}
H.~C. \"Ottinger.
\newblock {General projection operator formalism for the dynamics and thermodynamics of complex fluids}.
\newblock {\em Physical Review E}, 57:1416--1420, Feb 1998.

\bibitem{ottingerprojection2}
H.~C. \"Ottinger.
\newblock Derivation of a two-generator framework of nonequilibrium thermodynamics for quantum systems.
\newblock {\em Physical Review E}, 62:4720--4724, Oct 2000.

\bibitem{ottingerprojection3}
H.~C. Öttinger.
\newblock {Systematic Coarse Graining: “Four Lessons and A Caveat” from Nonequilibrium Statistical Mechanics}.
\newblock {\em MRS Bulletin}, 32(11):936–940, November 2007.

\bibitem{Mielke2011}
A.~Mielke.
\newblock {Formulation of thermoelastic dissipative material behavior using GENERIC}.
\newblock {\em Continuum Mechanics and Thermodynamics}, 23(3):233--256, May 2011.

\bibitem{GRMELA2012976}
M.~Grmela.
\newblock Fluctuations in extended mass-action-law dynamics.
\newblock {\em Physica D: Nonlinear Phenomena}, 241(10):976--986, 2012.

\bibitem{maxwell1867dynamical}
J.~C. Maxwell.
\newblock {IV. On the dynamical theory of gases}.
\newblock {\em Philosophical Transactions of the Royal Society of London}, 157:49--88, 1867.

\bibitem{cattaneo1958forme}
C.~Cattaneo.
\newblock {\em Sur une forme de l'{\'e}quation de la chaleur {\'e}liminant le paradoxe d'une propagation instantan{\'e}e}.
\newblock Comptes rendus hebdomadaires des s{\'e}ances de l'Acad{\'e}mie des sciences. Gauthier-Villars, 1958.

\bibitem{vernotte1958paradoxes}
P.~Vernotte.
\newblock Les paradoxes de la théorie continue de l’équation de la chaleur.
\newblock {\em Comptes Rendus de l’Académie des Sciences}, 246:3154--3155, 1958.

\bibitem{Grmela_2019}
M.~Grmela, M.~Pavelka, V.~Klika, B.-Y. Cao, and N.~Bendian.
\newblock {Entropy and Entropy Production in Multiscale Dynamics}.
\newblock {\em Journal of Non-Equilibrium Thermodynamics}, 44(3):217–233, May 2019.

\bibitem{nonfourier}
M.~Szücs, M.~Pavelka, R.~Kovács, T.~Fülöp, P.~Ván, and M.~Grmela.
\newblock {A Case Study of Non-Fourier Heat Conduction Using Internal Variables and GENERIC}.
\newblock {\em Journal of Non-Equilibrium Thermodynamics}, 47(1):31–60, September 2021.

\bibitem{AJJI2023133642}
A.~Ajji, J.~Chaouki, O.~Esen, M.~Grmela, V.~Klika, and M.~Pavelka.
\newblock On geometry of multiscale mass action law and its fluctuations.
\newblock {\em Physica D: Nonlinear Phenomena}, 445:133642, 2023.

\bibitem{grmela2025rheologicalmodelinggenericonsager}
M.~Grmela.
\newblock {Rheological modeling with GENERIC and with the Onsager principle}, 2025.
\newblock arXiv:2507.22144 [cond-mat.soft].

\bibitem{WAGNER2001177}
N.~J. Wagner.
\newblock {The Smoluchowski equation for colloidal suspensions developed and analyzed through the GENERIC formalism}.
\newblock {\em Journal of Non-Newtonian Fluid Mechanics}, 96(1):177--201, 2001.

\bibitem{twophasemodels}
M.~H\"utter.
\newblock {Thermodynamically consistent incorporation of the Schneider rate equations into two-phase models}.
\newblock {\em Phys. Rev. E}, 64:011209, Jun 2001.

\bibitem{hutter2012viscoplastic}
M.~Hütter and B.~Svendsen.
\newblock Thermodynamic model formulation for viscoplastic solids as general equations for non-equilibrium reversible–irreversible coupling.
\newblock {\em Continuum Mechanics and Thermodynamics}, 24(3):211--227, 2012.

\bibitem{liquidcrystal1}
J.~{Salmer{\'o}n-Hern{\'a}ndez}, P.~{Zubieta}, J.~{de Pablo}, and H.~{Öttinger}.
\newblock {Out-of-Equilibrium GENERIC Modeling Predicts Lyotropic Liquid Crystal Emulsions with Diffuse Interfaces}.
\newblock In {\em APS March Meeting Abstracts}, volume 2023 of {\em APS Meeting Abstracts}, page Z07.004, January 2023.

\bibitem{onsager19311}
L.~Onsager.
\newblock {Reciprocal Relations in Irreversible Processes. I.}
\newblock {\em Physical Review}, 37:405--426, Feb 1931.

\bibitem{onsager19312}
L.~Onsager.
\newblock {Reciprocal Relations in Irreversible Processes. II.}
\newblock {\em Physical Review}, 38:2265--2279, Dec 1931.

\bibitem{Kraaij_2020}
R.~C. Kraaij, A.~Lazarescu, C.~Maes, and M.~Peletier.
\newblock {Fluctuation symmetry leads to GENERIC equations with non-quadratic dissipation}.
\newblock {\em Stochastic Processes and their Applications}, 130(1):139–170, 2020.

\bibitem{quasisymmetry}
A.~Trautman.
\newblock {Noether equations and conservation laws}.
\newblock {\em Communications in Mathematical Physics}, 6, 1967.

\bibitem{MPR13}
A.~Mielke, M.~A. Peletier, and D.~R.~M. Renger.
\newblock {On the Relation between Gradient Flows and the Large-Deviation Principle, with Applications to Markov Chains and Diffusion}.
\newblock {\em Potential Analysis}, 41(4):1293–1327, 2014.

\bibitem{largemarkov}
R.~Kraaij.
\newblock {\em {Semigroup methods for large deviations of Markov processes}}.
\newblock PhD thesis, Delft University of Technology, 2016.

\bibitem{Maesring}
C.~Maes, K.~Netočný, and B.~Wynants.
\newblock Steady state statistics of driven diffusions.
\newblock {\em Physica A: Statistical Mechanics and its Applications}, 387(12):2675–2689, 2008.

\bibitem{mepp}
C.~Maes and K.~Netočný.
\newblock Minimum entropy production principle.
\newblock {\em Scholarpedia}, 8:9664, 2013.

\bibitem{Bruers_2007}
S.~Bruers, C.~Maes, and K.~Netočný.
\newblock {On the Validity of Entropy Production Principles for Linear Electrical Circuits}.
\newblock {\em Journal of Statistical Physics}, 129(4):725–740, 2007.

\bibitem{lagrangianmarkov}
C.~Maes, K.~Netočný, and B.~Wynants.
\newblock {On and beyond entropy production: the case of Markov jump processes}.
\newblock {\em Markov Processes and Related Fields}, 14:445--464, 2008.

\bibitem{laghamconnection}
A.~Deriglazov.
\newblock {\em {Classical Mechanics: Hamiltonian and Lagrangian Formalism}}.
\newblock Springer Cham, 2016.

\bibitem{nonconservativenoether}
D.S. {Djukic} and B.D. {Vujanovic}.
\newblock {Noether's theory in classical nonconservative mechanics}.
\newblock {\em Acta Mechanica}, 23(1-2):17--27, 1975.

\bibitem{Vittal1988}
V.~Vittal and A.~N. Michel.
\newblock A variational principle for nonconservative power systems.
\newblock {\em Circuits, Systems and Signal Processing}, 7(3):413--424, 1988.

\bibitem{goldstein2002classical}
H.~Goldstein, C.P. Poole, and J.L. Safko.
\newblock {\em {Classical Mechanics}}.
\newblock Addison Wesley, {Third} edition, 2002.

\bibitem{nonholonomic}
N.~A. Lemos.
\newblock {Complete inequivalence of nonholonomic and vakonomic mechanics}.
\newblock {\em Acta Mechanica}, 233(1):47--56, 2022.

\end{thebibliography}

\end{document}